\documentclass[reprint,aps,prresearch,amsmath,amssymb,superscriptaddress,floatfix]{revtex4-2}

\usepackage{graphicx}
\usepackage{dcolumn}
\usepackage{bm}
\usepackage{xcolor}
\usepackage[colorlinks=true,linkcolor=blue!55!black,citecolor=blue!55!black,urlcolor=blue!55!black]{hyperref}

\newcommand{\ket}[1]{\lvert #1\rangle}

\newcommand{\Lv}{\mathcal{L}}
\newcommand{\id}{\mathbb{I}}

\newcommand{\sgn}{\operatorname{sgn}}
\newcommand{\Imag}{\operatorname{Im}}
\newcommand{\Real}{\operatorname{Re}}
\newcommand{\rss}{\rho_{\mathrm{ss}}}
\newcommand{\nK}{\nu_{\mathrm{K}}}
\newcommand{\nB}{\nu_{\mathrm{B}}}
\newcommand{\nC}{\nu_{\mathrm{C}}}
\newcommand{\PZM}{P_{\mathrm{ZM}}}
\newcommand{\Plow}{P_{\mathrm{low}}}
\newcommand{\vac}{\ket{\mathrm{vac}}}

\makeatletter
\def\subsection{\@startsection{subsection}{2}{\z@}{.45cm \@plus.5ex \@minus .2ex}{.25cm}{\normalfont\small\bfseries\centering}}
\def\subsubsection{\@startsection{subsubsection}{3}{\z@}{.45cm \@plus.5ex \@minus .2ex}{.25cm}{\normalfont\small\itshape\centering}}
\makeatother

\begin{document}

\title{Phase-Tunable Liouvillian Topology: Dual Winding Structures and Skin Effects}

\author{Daksh K. Sambhare}
\email[]{d.sambhare@iitg.ac.in}
\author{Koustav Roy}
\email[]{koustav.roy@iitg.ac.in}
\author{Saurabh Basu}
\affiliation{Department of Physics, Indian Institute of Technology Guwahati, Guwahati 781039, Assam, India}


\begin{abstract}
Liouvillian skin effects are usually associated with dissipation whose rates favor unidirectional flow. In this work, we show that directional edge localization can emerge even when population exchange between sites is exactly balanced, with a single tunable phase parameter $\alpha$ deciding their direction and a dual winding structure deciding their nature. Although this number-conserving bath is known for eigenstate selection, we demonstrate here that its phase controls a steady current, which Hamiltonian hopping converts to directional edge localization. Further, we study the spectral winding structure by independently twisting the left (ket) and right (bra) vector indices of the density matrix, revealing two distinct windings whose relationship is dictated by an antiunitary mirror symmetry. Nearest neighbor jump operators break this symmetry: both windings coincide, giving rise to a common spectral winding $W_0(\alpha)=\pm1$, producing a unipolar skin effect in which the density accumulates at a specific edge. However, next nearest neighbor jump preserve it. The two windings are forced to be opposite, and the ket and bra are pushed to opposite edges, giving a bipolar skin effect of coherences present in eigenmodes, without edge localization of the density in the steady state. Remarkably, in a dimerized chain the phase-controlled density skin effect gives the steady-state zero modes a size-independent weight at the skin edge, accompanied by a kink at the topological transition. In addition, the relaxation time grows linearly with the system size, however at specific phases $\alpha$ where the jump operators are rendered Hermitian, edge-localized eigenmodes with skin character coexist with diffusive relaxation.
\end{abstract}
\maketitle

\section{Introduction}
\label{sec:intro}

Open quantum systems coupled to Markovian baths evolve under the Lindblad master equation, whose generator, the Liouvillian superoperator, is non-Hermitian~\cite{Lindblad1976,Gorini1976,Breuer2002}. The complex Liouvillian spectrum can exhibit point-gap topology~\cite{Gong2018,Kawabata2019,Okuma2020,Zhang2020,Borgnia2020,Chaduteau2026,Jha2026}, with nonzero spectral winding giving rise to the Liouvillian skin effect (LSE) under open boundary conditions, characterized by the accumulation of a macroscopic number of eigenmodes at a boundary~\cite{Song2019,Haga2021,Yang2022,Wang2023,Hamanaka2023,Mao2024,Feng2024,Ekman2024,Long2026,RoyShahab2025}. The LSE also reshapes dissipative dynamics, most strikingly through relaxation times that diverge with system size despite an open Liouvillian gap~\cite{Haga2021}.

In the mechanisms studied so far, the direction of the LSE is intrinsically encoded in the dissipative processes. Asymmetric incoherent hopping ($\gamma_R\neq\gamma_L$) induces preferential particle transport toward one boundary~\cite{Haga2021,Wang2023}, while unidirectional transfer between the sublattices of a chiral chain imprints the band winding onto the Liouvillian spectrum~\cite{Long2026}. Gain and loss produce a nonreciprocal damping matrix~\cite{Song2019} and two-body loss yields an LSE only in combination with an inversion-breaking Hamiltonian~\cite{Hamanaka2023}. In each case the bath itself distinguishes left from right. Skin effects without asymmetric hopping are known for short-time approximated effective non-Hermitian Hamiltonians, where on site gain and loss act on bands with spin-momentum locking~\cite{OkumaSato2023}, however for Liouvillians the push has always been supplied by the bath. Could a bath with exactly balanced population exchange still produce a skin effect, and if yes then what selects its direction.

A natural setting for this question is a bath that has been studied extensively for a different purpose. The following number conserving jump operator $O_j=(c_j^\dagger+e^{i\alpha}c_{j+l}^\dagger)(c_j-e^{i\alpha}c_{j+l})$ removes a particle from one superposition of the sites $j$ and $j+l$ and returns it into another. For $\alpha=0$, the jump operator corresponds to reservoir-engineering schemes that lock the relative phase between neighboring sites~\cite{Diehl2008,Kraus2008,Diehl2010} and enable dissipative preparation of topological states~\cite{Diehl2011,Bardyn2013}, while its effects in interacting chains were investigated in Ref.~\cite{Bonnes2014}. With a tunable phase it acts as a phase-selective filter in energy: it drives quasiperiodic chains into extended or localized steady states~\cite{Liu2024}, prepares the edge states of a topological chain when it acts within one sublattice~\cite{Peng2024a}, and reshapes the relaxation and localization caused by boundary dissipation~\cite{Peng2024b} and Mpemba-type effects~\cite{Liu2025,Strachan2025}. In all these studies, the phase determines the occupation of eigenstates. However, the real-space consequences of this phase remain unexplored from the perspective of Liouvillian topology. Phase-dependent jumps are known to produce Liouvillian skin effects for linear loss and gain baths~\cite{Song2019}. To our knowledge, this question has not been examined for a number-conserving bath whose population exchange is exactly balanced and whose phase acts solely on the coherence between two sites.

Characterizing such an effect raises a second question. The Liouvillian acts on the density matrix from two sides at once, and a boundary twist can be applied to either side on its own. This method gives two separate winding numbers instead of one, windings of the Liouvillian's spectrum, and their effects on the density and on the coherences may be distinct. The invariants used for the LSE so far give only one such number. Either the winding in the quasimomentum transfer $K$, where the two sides left(ket) and right(bra) are equally affected~\cite{Long2026} or a twist applied to one side alone~\cite{Hamanaka2023}; where two windings have been compared, they belong to a Lindbladian and its postselected non-Hermitian effective Hamiltonian~\cite{Chaduteau2025}. Twisting both sides independently defines instead a pair of windings $(\nK,\nB)$.


\begin{figure}[t]
\includegraphics[width=\columnwidth]{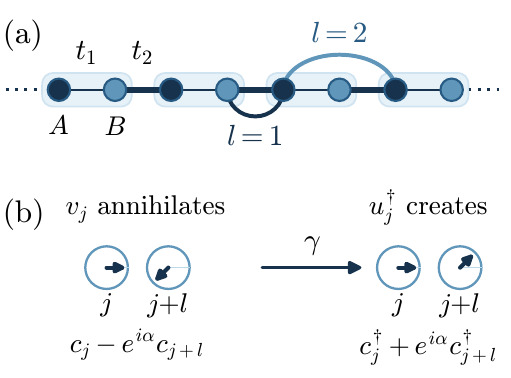}
\caption{(a)~Hoppings $t_1$ (thin) and $t_2$ (thick); jumps act on bonds of range $l=1$ (A--B) or $l=2$ (A--A, B--B). (b)~A jump at rate $\gamma$ empties $v_j$ and refills $u_j$; arrows give the phase of each amplitude relative to site $j$.}
\label{fig:model}
\end{figure}

In this work we address both questions for a dimerized chain~\cite{Su1979,Asboth2016} coupled to bond jumps of range $l=1$ and $l=2$. Several interesting features emerge. First, every jump exchanges population symmetrically on their own, but their two parts interfere through the bond coherences. Together with the coherent hopping, this interference drives a steady current around a periodic boundary, which an open boundary turns into edge localization. Second, for nearest-neighbor bonds, this directional push is captured by a Liouvillian winding $W_0(\alpha)=\pm1$, which belongs to the bath and is therefore the same in the topological and trivial phases of $H$, unlike in Ref.~\cite{Long2026}. Next, an antiunitary mirror-chiral symmetry decides whether $\nu_{\rm K}$ and $\nu_{\rm B}$ must move together. Nearest-neighbor jumps break it, and the density accumulates. Next-nearest-neighbor jumps preserve it, the ket and bra are forced to opposite edges instead, a bipolar skin effect. Furthermore, where the density accumulates, the zero-mode weight switches on with a kink at topological transition. Finally, the accumulation also makes the relaxation time grow linearly with system size~\cite{Haga2021}, while at the Hermitian points relaxation stays diffusive even though the eigenmodes are edge localized. More broadly, our results demonstrate that directional dissipative transport can emerge from interference between coherent and incoherent processes, without requiring asymmetric rates. They further reveal that two indices of the density matrix can exhibit distinct point-gap windings~\cite{Okuma2020,Zhang2020,Hamanaka2023}, and that symmetry-protected bipolar localization~\cite{Okuma2020,RoyGogoi2025,Roy2026} has a Liouvillian counterpart manifested in the coherences.

The remainder of this paper is organized as follows: Section~\ref{sec:model} introduces the model, the selection of eigenstates (Sec.~\ref{sec:energy}), and the skin effect with its dual windings (Sec.~\ref{sec:lse}). Sections~\ref{sec:consequences} and \ref{sec:relax} discuss the current, the zero modes, and the relaxation, Sec.~\ref{sec:experiment} a realization, and Sec.~\ref{sec:summary} the outlook; derivations can be found in the Supplementary Material.

%
\section{Model and the dissipator}
\label{sec:model}

We consider a particle hopping on a one-dimensional lattice of $N$ sites ($N$ even), labeled $i=0,\dots,N-1$, with sublattice A on even and B on odd sites, described by the Hamiltonian
\begin{equation}
H=\sum_{i=0}^{N-2}t_i\big(c_i^\dagger c_{i+1}+\mathrm{h.c.}\big),\qquad
t_i=\begin{cases}t_1, & i\ \text{even},\\ t_2, & i\ \text{odd},\end{cases}
\label{eq:H}
\end{equation}
where $c_i^\dagger$ and $c_i$ create and annihilate a particle on site $i$. For $t_1<t_2$ the off-diagonal Bloch block $h(k)=t_1+t_2e^{-ik}$ winds once around the origin, binding one zero mode to each edge (topological phase) and for $t_1>t_2$ it does not wind (trivial phase)~\cite{Su1979,Asboth2016,RoyBasu2023}. Our results use only that $H$ is real, bipartite, and mirror symmetric namely,
\begin{equation}
\Gamma H\Gamma^{-1}=-H,\qquad PHP^{-1}=H,\qquad H^*=H,
\label{eq:Hsym}
\end{equation}
where the sublattice sign $\Gamma$ maps to $c_i\to(-1)^ic_i$ and the selected bond mirror $P$ maps to $c_i\to c_{N-1-i}$. According to the Born-Markov approximation~\cite{Breuer2002,Daley2014}, the density matrix obeys the Lindblad master equation \cite{Lindblad1976,Gorini1976}
\begin{equation}
\dot\rho=\Lv_\alpha\rho=-i[H,\rho]+\gamma\sum_j\Big(O_j\rho O_j^\dagger-\tfrac12\big\{O_j^\dagger O_j,\rho\big\}\Big),
\label{eq:lindblad}
\end{equation}
with the number-conserving bond jumps
\begin{equation}
O_j=\big(c_j^\dagger+z\,c_{j+l}^\dagger\big)\big(c_j-z\,c_{j+l}\big),\qquad z=e^{i\alpha},
\label{eq:O}
\end{equation}
acting on every bond $(j,j+l)$, $j=0,\dots,N-1-l$. All bonds are oriented along the chain, $j\to j+l$; this orientation, together with the phase $\alpha$ distinguishes the two ends of density matrices. We study $l=1$, where each bond joins sites of different sublattices, and $l=2$, where both sites of a bond lie on one sublattice [Fig.~\ref{fig:model}(a)]. We restrict our analysis to single-particle sector, since $O_j$ conserves particle number.

\emph{Balanced Current -- }
Each jump factorizes into one annihilation and one creation parts of bond,
\begin{equation}
O_j=u_j^\dagger v_j,\qquad u_j^\dagger=c_j^\dagger+z\,c_{j+l}^\dagger,\qquad v_j=c_j-z\,c_{j+l}.
\label{eq:rank1}
\end{equation}
So a jump removes a particle from $v_j$ and returns it into $u_j$ [Fig.~\ref{fig:model}(b)], with $\{u_j,u_j^\dagger\}=\{v_j,v_j^\dagger\}=2$.
Expanding Eq.~\eqref{eq:O} in the site operators separates the jump into an on-site and a transfer part,
\begin{equation}
\begin{aligned}
O_j&=O_j^{D}+O_j^{T},\qquad O_j^{D}=n_j-z^2\,n_{j+l},\\
O_j^{T}&=z\big(c_{j+l}^\dagger c_j-c_j^\dagger c_{j+l}\big),
\end{aligned}
\label{eq:DT}
\end{equation}
with $n_i=c_i^\dagger c_i$. In $O_j^{T}$, the two hopping amplitudes differ only in sign while retaining equal magnitudes for an arbitrary $\alpha$. This symmetry is preserved in the population dynamics of $n_j$, for which the bath induces only symmetric particle exchange across an individual bond, irrespective of the coherences (see Sec.~\ref{supp:sym} of the Supplementary Material),
\begin{equation}
\dot n_j\big|_{\text{bond}}=\gamma\,(n_{j+l}-n_j)=-\dot n_{j+l}\big|_{\text{bond}}.
\label{eq:flatten}
\end{equation}
By itself, the bath thus only flattens the density. This is the essential difference from incoherent hopping $\sqrt{\gamma_R}\,c_{j+1}^\dagger c_j$ and $\sqrt{\gamma_L}\,c_j^\dagger c_{j+1}$, where the ratio $\gamma_R/\gamma_L$ usually sets the direction~\cite{Haga2021}.

\begin{figure}[t]
\includegraphics[width=\columnwidth]{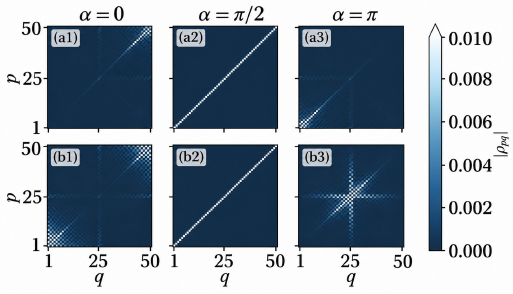}
\caption{The bath phase selects a band for $l=1$ and an energy magnitude for $l=2$. Steady-state $|\rho_{pq}|$ in the eigenbasis of $H$ (ascending energy; zero modes at $p,q=25,26$), $t_1=1.0$, $t_2=1.5$, $\gamma=1$, $N=50$. (a1)--(a3)~$l=1$ and (b1)--(b3)~$l=2$ at $\alpha=0,\pi/2,\pi$.}
\label{fig:energy}
\end{figure}

\emph{Hermitian points, interference, and symmetry --- }A bath that only flattens populations might be expected to leave the fully mixed state $\rho=\id/N$ untouched. However, it does not. Evaluating the dissipator on $\id$ shows that it pumps energy into every bond,
\begin{equation}
\Lv_\alpha(\id)=2\gamma\,(z+z^*)\sum_j\big(c_j^\dagger c_{j+l}+\mathrm{h.c.}\big),
\label{eq:unital}
\end{equation}
and this pump switches off only at the Hermitian points $\alpha=\pi/2+n\pi$, where $O_j^\dagger=O_j$ and the identity is left invariant, $\rss=\id/N$ (Sec.~\ref{supp:sym}).
Taken alone, each of $O_j^{D}$ and $O_j^{T}$ generates unital dynamics with $\rss=\id/N$ for
every $H$ and $\alpha$, so every $\alpha$-dependent feature of $\rss$ stems from their
interference,
\begin{equation}
\begin{aligned}
\mathcal{D}\big[O^{D}+O^{T}\big]&=\mathcal{D}\big[O^{D}\big]+\mathcal{D}\big[O^{T}\big]+\mathcal{X},\\
\mathcal{X}(\rho)&=O^{D}\rho\, O^{T\dagger}+O^{T}\rho\, O^{D\dagger}\\
&\quad-\tfrac12\big\{O^{D\dagger}O^{T}+O^{T\dagger}O^{D},\rho\big\}.
\end{aligned}
\label{eq:interference}
\end{equation}
The interference requires coherent hopping to act on and for $H=0$ the steady-state density is symmetric at every $\alpha$ (Sec.~\ref{supp:sym}).

We now build the symmetry that governs this by reflecting the chain about its center, applying the sublattice sign, and complex-conjugate in the site basis, $\mathcal{A}(\rho)=P\Gamma\rho^{*}\Gamma P$. The coherent dynamics remains invariant for any $H$ satisfying Eq.~\eqref{eq:Hsym}, since both conjugation and the sublattice transformation reverses the hopping, so the two cancel. Mirror reflection and complex conjugation each send $\alpha\to-\alpha$, so together they leave the phase of the jumps unchanged, and only the sublattice sign still acts on it. There its effect depends on where the bond sits: for $l=2$ both sites of a bond share one sublattice and $\Gamma$ leaves the jump alone, so
\begin{equation}
\mathcal{A}\Lv_\alpha\mathcal{A}^{-1}=\Lv_\alpha\qquad(l=2),
\end{equation}
a symmetry at every $\alpha$. For $l=1$ the two sites of a bond lie on opposite sublattices, and $\Gamma$ shifts the phase instead, $\Gamma O_j(\alpha)\Gamma^{-1}=O_j(\alpha+\pi)$ (Sec.~\ref{supp:sym}), so that
\begin{equation}
\mathcal{A}\Lv_\alpha\mathcal{A}^{-1}=\Lv_{\alpha+\pi}\qquad(l=1),
\label{eq:sym}
\end{equation}
which relates two baths that push toward opposite edges rather than leaving $\Lv_\alpha$ fixed. Since mirror reflection alone already maps $\Lv_\alpha$ onto $\Lv_{-\alpha}$, the phase $\alpha$ is the only parameter in the model that breaks inversion, and hence the only one that can select which edge the bath pushes toward.

\subsection{Dissipative selection in the energy basis}
\label{sec:energy}

The bath acts as a phase-selective pump in energy, a property that underlies its use for eigenstate selection~\cite{Liu2024,Peng2024a}. A jump on the bond $j$ removes the particle from the orbital $v_j$, with amplitudes $(1,-z^*)$ on the sites $(j,j+l)$, and returns it into $u_j$, with amplitudes $(1,z)$. A single particle state $\sum_i\psi_ic_i^\dagger\vac$ is emptied with an amplitude -
\begin{equation}
v_j\sum_i\psi_ic_i^\dagger\vac=\big(\psi_j-z\,\psi_{j+l}\big)\vac
\label{eq:overlap}
\end{equation}
The two orbitals coincide at the Hermitian points, where a jump returns the particle to the orbital it came from. Further, they are orthogonal at $\alpha=0,\pi$, where the transfer between them is complete and in between it is partial.

For $l=1$ the two orbitals carry the bond energies $\mp t_j\cos\alpha$, so every jump raises the energy by $2t_j\cos\alpha$, the pump of Eq.~\eqref{eq:unital}. The bath therefore fills the upper band for $\cos\alpha>0$ and the lower band for $\cos\alpha<0$, most strongly at $\alpha=0$ and $\pi$, where the refilled orbital is the in-phase or the out-of-phase combination of two neighbors.
\begin{figure*}[t]
\includegraphics[width=\textwidth]{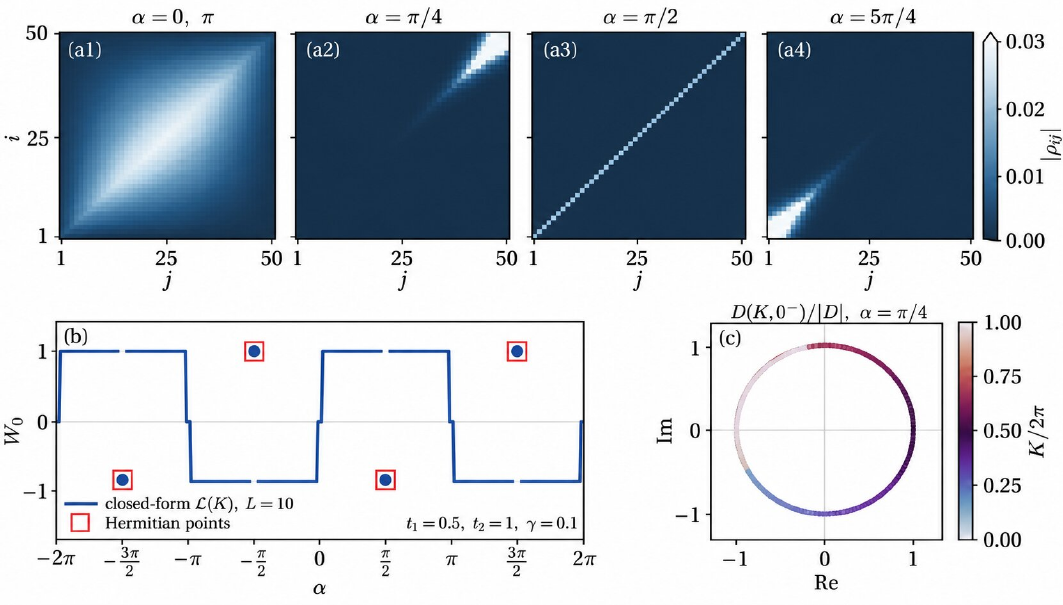}
\caption{The bath phase sets the direction of the $l=1$ density skin effect, and the $K$-sector winding counts it. (a1)--(a4)~Steady-state $|\rho_{ij}|$ in the site basis (sites labeled $1,\dots,N$; $N=50$, $t_1=1.5$, $t_2=2.5$, $\gamma=1$) at $\alpha=0$ and $\pi$ (identical), $\pi/4$, $\pi/2$, and $5\pi/4$. (b)~$W_0(\alpha)$ from Eqs.~\eqref{eq:LK} and \eqref{eq:W0} for $L=10$ unit cells, $t_1=0.5$, $t_2=1$, $\gamma=0.1$. Squares mark the Hermitian points $\alpha=\pi/2+n\pi$, where the evaluated value (dots) is not an invariant. (c)~Normalized determinant $D(K,0^-)/|D|$ over the Brillouin zone at $\alpha=\pi/4$, colored by $K/2\pi$, encircling the origin once.}
\label{fig:skin}
\end{figure*}
 For $l=2$ the jump compares amplitudes two sites apart on one sublattice. At $\alpha=0$ it spares states with $\psi_{j+2}=\psi_j$, which on a ring are the outer band edges $E=\pm(t_1+t_2)$, and at $\alpha=\pi$ states with $\psi_{j+2}=-\psi_j$, the gap edges $E=\pm|t_2-t_1|$; on the open chain the steady state favors the states closest to these patterns. The zero modes share the sign pattern favored at $\alpha=\pi$ but not its magnitude, and are emptied only weakly (Sec.~\ref{supp:ss}).

The symmetry~\eqref{eq:sym} makes this useful for any $H$ obeying Eq.~\eqref{eq:Hsym}. Let $d_E$ denote the eigenmode of $H$ at energy $E$, and $p_E=\langle d_E^\dagger d_E\rangle$ its steady-state population. Since the product of the mirror and the sublattice sign (we call $P\Gamma$), maps $d_E$ onto $\pm d_{-E}$, these populations obey (Sec.~\ref{supp:sym})
\begin{equation}
\begin{cases}
p_E(\alpha)=p_{-E}(\alpha), & l=2,\\[2pt]
p_E(\alpha+\pi)=p_{-E}(\alpha), & l=1.
\end{cases}
\label{eq:pE}
\end{equation}
For $l=2$ both bands are always equally filled and $\alpha$ selects only  the magnitude, i.e., $|E|$; for $l=1$ the shift $\alpha\to\alpha+\pi$ exchanges the bands, as it later reverses the skin effect. Figure~\ref{fig:energy} shows both. For $l=1$ the weight appears in the upper (lower) band at $\alpha=0$ ($\pi$), and is equipartitioned at Hermitian point. For $l=2$ it gathers at the outer band edges at $\alpha=0$ and around $E=0$ at $\alpha=\pi$, where the in-gap zero-mode pair holds well above its equipartition weight.~\cite{Peng2024a}.

\subsection{Liouvillian skin effect}
\label{sec:lse}

Equation~\eqref{eq:flatten} excludes any directional push from the populations alone. In this section we show that the push nevertheless appears, first in the density for $l=1$, where a winding in the quasimomentum transfer fixes its direction (Sec.~\ref{sec:density}), and then in the two indices of $\rho$ separately, where $l=1$ and $l=2$ behave in opposite ways (Sec.~\ref{sec:twoindex}).

\subsubsection{Density skin effect for \texorpdfstring{$l=1$}{l=1} and its spectral winding}
\label{sec:density}

On a translationally invariant ring the Liouvillian couples the blocks $\rho(k,k')$ only within sectors of fixed quasimomentum transfer $K=k'-k$, which plays the role of the Bloch momentum of an effective one-dimensional Liouvillian, while $k$ labels its internal degrees of freedom~\cite{Long2026}. For $l=1$ each sector takes the closed form
\begin{equation}
\begin{aligned}
\Lv(K)={}&\mathrm{diag}\big[\Lv(K,k_1),\dots,\Lv(K,k_L)\big]\\
&+N_L\otimes M_0^{\mathrm{intra}}+S\big[N_L\otimes \tilde M_0(K)\big]S^{-1},
\end{aligned}
\label{eq:LK}
\end{equation}
where $\Lv(K,k)$ acts within the block $\rho(k,k+K)$, $N_L$ is the $L\times L$ all-ones matrix ($L=N/2$), and $M_0^{\mathrm{intra}}$ and $\tilde M_0(K)$ carry the jumps of the intracell and intercell bonds (Sec.~\ref{supp:LKk}). Its determinant $D(K,\lambda)=\det[\Lv(K)-\lambda]$ reduces to blocks of fixed size and is evaluated in closed form at every $\alpha$.

The steady state sits at $\lambda=0^-$, and the topology of the band that supports it is read
from the spectral winding about a reference just below,
\begin{equation}
W_0=\frac{1}{2\pi i}\oint dK\,\partial_K\ln \det[\Lv(K)-0^-].
\label{eq:W0}
\end{equation}
Nonzero spectral winding implies skin localization~\cite{Okuma2020,Zhang2020,Borgnia2020}, so the steady state should pile up along the diagonal of $\rss$ whenever $W_0\neq0$, at the right edge for $W_0>0$; $W_0$ counts how often $D(K,0^-)/|D|$ encircles the origin as $K$ sweeps the Brillouin zone [Fig.~\ref{fig:skin}(c)]. Because $\alpha$ is the only element of the model that breaks inversion and Eq.~\eqref{eq:sym} ties it to $\alpha+\pi$,
\begin{equation}
W_0(-\alpha)=-W_0(\alpha),\qquad W_0(\alpha+\pi)=-W_0(\alpha).
\label{eq:W0sym}
\end{equation}
Evaluating Eq.~\eqref{eq:W0} gives
\begin{equation}
W_0(\alpha)=\begin{cases}
+1, & 0<\alpha<\pi,\\
-1, & \pi<\alpha<2\pi,\\
\ \ 0, & \alpha=0,\ \pi,
\end{cases}
\qquad \alpha\neq\tfrac{\pi}{2}+n\pi,
\label{eq:W0piece}
\end{equation}
[Fig.~\ref{fig:skin}(b)] depicts within each half-period no smooth change of $\alpha$ can reverse
the push. At the Hermitian points the dynamics is unital and the steady-state branch touches
the reference point, so the value returned there [dots in Fig.~\ref{fig:skin}(b)] carries no
protected meaning, consistent with the featureless steady state.

Figure~\ref{fig:skin}(a) confirms the expected localization. The steady state piles up at
the right edge for $0<\alpha<\pi$ and at the left edge for $\pi<\alpha<2\pi$, over a decay
length of about two unit cells that does not change with $N$, which identifies a bulk skin
effect rather than the tail of an edge state. The profile is mirror symmetric at
$\alpha=0,\pi$, is mirror reflected under $\alpha\to\alpha+\pi$, and is flat at $\alpha=\pi/2+n\pi$ (Sec.~\ref{supp:sym}), following $W_0$ throughout. The push therefore needs the interference term~\eqref{eq:interference} and coherent hopping for it to
act on, together with broken inversion ($\alpha\neq0,\pi$) and non-unitality
($\alpha\neq\pi/2+n\pi$).

$W_0$ counts the displacement of the ket and the bra together, as a single density. Since
$\rho$ carries two indices, we next twist them separately.
\begin{figure}[t]
\includegraphics[width=\columnwidth]{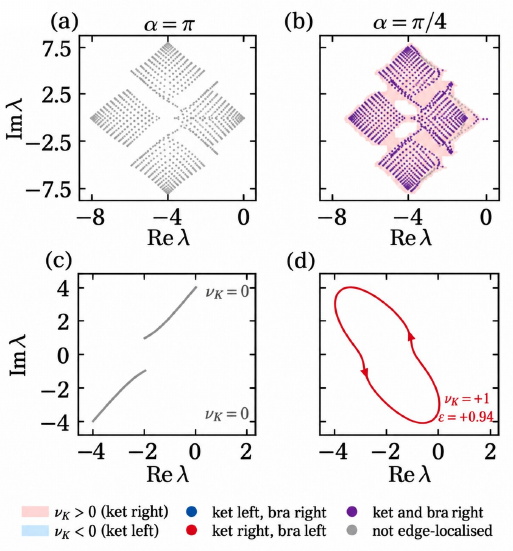}
\caption{For $l=1$ the bath pushes ket and bra to the same edge, and the skin effect appears in the density. Topological phase ($t_1=1.5$, $t_2=2.5$), $\gamma=1$ and ($N=30$). (a)~$\alpha=\pi$: no edge-localized modes. (b)~$\alpha=\pi/4$: every edge-localized mode has ket and bra at the right edge (purple), on a single region with $\nK>0$. (c),(d)~Vacuum-sector loops: two arcs with $\nK=0$ at $\alpha=\pi$; one loop with $\nK=+1$ at $\alpha=\pi/4$, whose open-chain modes sit at the right edge.}
\label{fig:l1}
\end{figure}
\subsubsection{Density versus coherence: windings of ket and bra}
\label{sec:twoindex}

The closed form~\eqref{eq:LK} is specific to the $l=1$ bonds, and $W_0$ in any case counts
only the joint displacement of the two indices. A twisted ring resolves them and requires no
unit cell~\cite{Hamanaka2023}: every term of $\Lv$ that moves a particle forward across one
chosen bond, $c_q^\dagger c_p$ with $p=N-1$ and $q=0$, is multiplied by $e^{i\theta}$. The
twist is a flux through the ring, and a nonzero winding of the spectrum under it means that
the generator carries the twisted index around the ring in one direction; on an open chain the
same transport piles that index at one end. Because the particle number is conserved, the
operators acting on the ket and those acting on the bra of $\rho$ can be twisted separately
(Sec.~\ref{supp:twist}), which defines a ket winding $\nK$, a bra winding $\nB$, and, twisting
both together, a density winding $\nC$,
\begin{equation}
\nu_{\mathrm{X}}(\Lambda)=\frac{1}{2\pi i}\int_0^{2\pi}d\theta\,\partial_\theta
\ln\det\big[\Lv_{\mathrm{X}}(\theta)-\Lambda\big],
\label{eq:nu}
\end{equation}
where $\mathrm{X}=\mathrm{K},\mathrm{B},\mathrm{C}$ labels the twisted index and positive
values correspond to the right edge.

One relation holds for any number-conserving Lindbladian (Sec.~\ref{supp:sym}). Hermiticity
preservation means that conjugating $\rho$ exchanges the two twists with reversed sign, so
\begin{equation}
\nB(\Lambda^*)=\nK(\Lambda).
\label{eq:Jrule}
\end{equation}
On the real axis, the steady state and fast-relaxing modes have equal windings, so ket and bra move together, localizing the density. Opposite motion occurs at complex-conjugate eigenvalues, where the ket and bra windings are opposite.

\begin{figure*}[t]
\includegraphics[width=\textwidth]{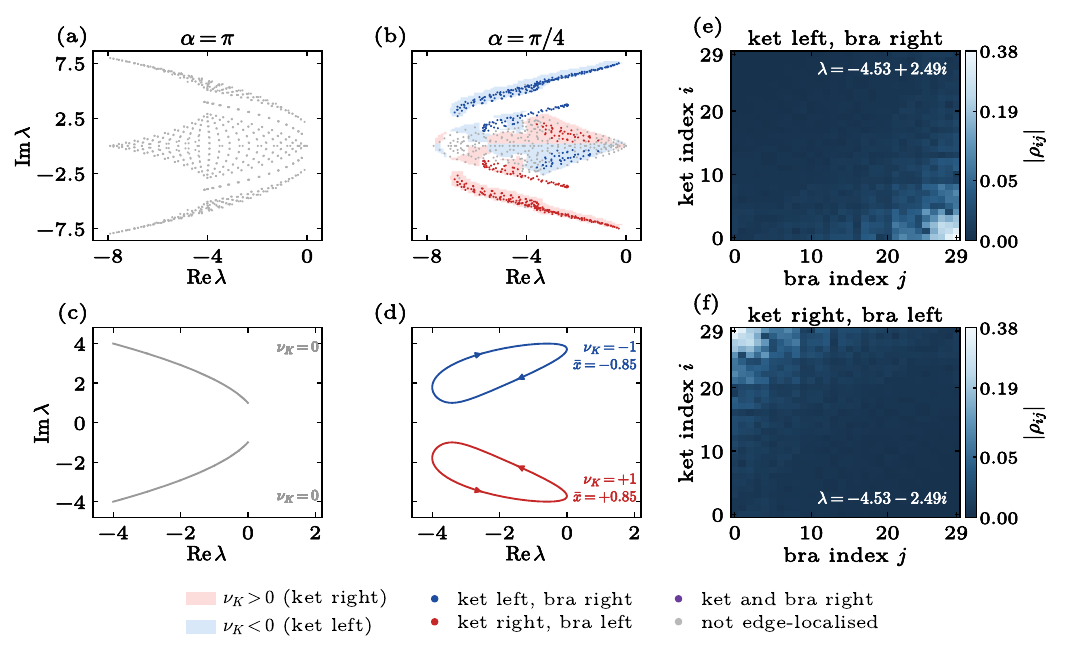}
\caption{For $l=2$ the bath pushes ket and bra to opposite edges: a bipolar skin effect of coherences. Topological phase ($t_1=1.5$, $t_2=2.5$), $\gamma=1$. (a),(b)~Spectra of the density block on the open chain ($N=30$) at $\alpha=\pi$ and $\pi/4$. Points are colored by the corner in which the eigenmode localizes (blue: ket left, bra right; red: ket right, bra left; gray: not edge localized). Shading is the sign of the ket winding $\nK$ on a 12-site twisted ring (red: $\nK>0$, ket pushed right; blue: $\nK<0$, ket pushed left). (c),(d)~Loops of the vacuum sector on a twisted ring at $\alpha=\pi$ and $\pi/4$; arrows give the sense of rotation and $\bar x$ the mean position of the corresponding open-chain modes. (e),(f)~$|\rho_{ij}|$ of the conjugate eigenmodes at $\lambda=-4.53\pm2.49i$: ket at the left and bra at the right edge, and the reverse.}
\label{fig:l2}
\end{figure*}

For $l=1$ a twist of $2\pi$ on either index moves every $K$ sector of Eq.~\eqref{eq:LK} by one
grid step, so that at the steady-state reference (Sec.~\ref{supp:LKk})
\begin{equation}
\nK(0^-)=\nB(0^-)=W_0,\qquad \nC(0^-)=2W_0 .
\label{eq:nuW0}
\end{equation}
The closed-form $W_0$ is thus the winding of the density, with ket and bra pushed together.
Remarkably, this invariant belongs to the bath rather than to the band topology of the bare
Hamiltonian. With jumps on every bond, a shift of the unit cell by one site maps the ring
Liouvillian for $(t_1,t_2)$ onto that for $(t_2,t_1)$, so that $W_0$ and the bulk skin effect
are identical in the topological and trivial phases; the two open chains differ only at their
ends. This should be contrasted with Ref.~\cite{Long2026}, where jumps that anticommute with
the sublattice operator let the band winding of the Hamiltonian set $W_0$, while jumps that
commute with it leave $W_0$ to the dissipative channels alone. The $l=1$ bond jump belongs to
neither class: its on-site part commutes with $\Gamma$ and its transfer part anticommutes, and
it is precisely the interference of the two that produces the skin effect.

The same push is already visible in the simplest sector of the dynamics. The coherence between
one particle and the empty chain, $\rho_{i,\mathrm{vac}}=\langle\mathrm{vac}|c_i\rho\vac$,
forms a closed block of the full Liouvillian generated by
$G=-iH-(\gamma/2)\sum_jO_j^\dagger O_j$; the jump term is present but acts on nothing in this
block (Sec.~\ref{supp:vacuum}). Twisting its single index yields one closed loop per band.
Figure~\ref{fig:l1} confirms the $l=1$ picture. At $\alpha=\pi/4$ every edge-localized
eigenmode of the density block has ket and bra at the same edge, on a single region with
$\nK>0$, and the vacuum sector shows one loop with $\nK=+1$ whose open-chain modes sit at the
right edge. At $\alpha=\pi$, where every winding vanishes, the loops collapse and no
eigenmode is edge localized.

Chirality changes the constraint. For a Hamiltonian satisfying Eq.~\eqref{eq:Hsym} and jumps
confined to one sublattice, as for $l=2$, the symmetry $\mathcal{A}$ holds at fixed $\alpha$
and leaves every twist unchanged. Combined with Hermitian conjugation it becomes the linear map
$\rho\mapsto P\Gamma\rho^{T}\Gamma P$, which sends the ket twist onto the reversed bra twist
and the density twist onto itself reversed, so that at every reference point
\begin{equation}
\nB(\Lambda)=-\nK(\Lambda),\qquad \nC(\Lambda)=0\qquad(l=2).
\label{eq:Crule}
\end{equation}
The content of Eq.~\eqref{eq:Crule} is read off from the two factors of $\mathcal{A}$:
conjugation exchanges the roles of ket and bra, and the mirror exchanges the two ends, so
whatever the bath does to the ket at one edge it does to the bra at the other. Together with
Eq.~\eqref{eq:Jrule} this gives $\nK=\nB=0$ on the real axis: the $l=2$ bath has no winding at
the steady-state reference and hence no directional skin effect of the density, and the
opposite windings it does allow are pushed off the axis. Equation~\eqref{eq:Crule} also
identifies where the asymmetry between the two indices originates. The sublattice sign treats
ket and bra alike; what distinguishes them is the preservation of Hermiticity,
Eq.~\eqref{eq:Jrule}, while $\Gamma$ only undoes the sign change of $H$ under conjugation,
which is possible at fixed $\alpha$ only when each jump preserves chirality. For $l=1$ the same
operation shifts $\alpha$ by $\pi$, so it relates two baths that push opposite ways and leaves
the density free to move.

The relation~\eqref{eq:Crule} has a direct consequence for the open-chain eigenmodes, each of
which is a table over the ket index $i$ and the bra index $j$. Since the ket winding at an
eigenvalue $\lambda$ predicts the edge at which the ket of the corresponding eigenmode
localizes, and likewise for the bra, Eq.~\eqref{eq:Crule} allows only crossed corners for
$l=2$: the ket at one end and the bra at the other. Each such eigenmode is a coherence between
the two ends rather than a population at either, its partner at $\lambda^*$ has ket and bra
exchanged, and its diagonal remains delocalized. The arrangement resembles the $\mathbb{Z}_2$ skin effect, in which a symmetry places the two members of a partner pair at opposite ends~\cite{Okuma2020} and here the two partners are the ket and the bra of a single eigenmode, paired by $\mathcal{A}$ together with Hermiticity preservation.

Figure~\ref{fig:l2} confirms this picture. At $\alpha=\pi/4$ [Fig.~\ref{fig:l2}(b)] all
edge-localized eigenmodes occupy the crossed corners, in equal numbers, and the ket side
follows the sign of $\nK$ at the eigenvalue. Figures~\ref{fig:l2}(e) and \ref{fig:l2}(f) display a conjugate pair
with ket and bra at opposite ends. At $\alpha=\pi$ [Fig.~\ref{fig:l2}(a)], where all windings
vanish, no eigenmode is edge localized. These are genuine skin modes: a weak coupling between
the two ends shifts their eigenvalues by an amount that grows exponentially with $N$, and their
left eigenvectors occupy the mirror corner (Sec.~\ref{supp:vacuum}). In the vacuum sector the
two band loops wind in opposite directions for $0<\alpha<\pi$ [Fig.~\ref{fig:l2}(d)], with
modes at the left and right edges, and both collapse into arcs at
$\alpha=\pi$ [Fig.~\ref{fig:l2}(c)], whose signs the outer arcs in Fig.~\ref{fig:l2}(b)
inherit. The vacuum loops need less symmetry than Eq.~\eqref{eq:Crule} (Sec.~\ref{supp:vacuum}).

The full dynamics, jumps included, does not transfer this bipolar structure to the steady
state, which sits on the real axis where Eq.~\eqref{eq:Crule} forces both windings to vanish.
Each jump refills a bond orbital and erases the band polarization built up by the push, so that
the end-to-end coherence is absent from $\rss$, and its residual imprint, a mirror-symmetric
enhancement of short-range coherence at both edges, fades roughly as $N^{-1/2}$. The bipolar
skin effect is thus a property of the decaying modes and of transient coherence: a coherence
prepared in the bulk spreads to both ends equally before it decays (Sec.~\ref{supp:vacuum}).

\section{Results and Analysis}
\label{sec:consequences}

The skin effect determines the spatial accumulation of the steady-state weight. We next investigate whether the zero modes of the bare Hamiltonian inherit this accumulation (Sec.~\ref{sec:zm}) and how it is maintained against the bath-induced tendency toward a uniform distribution (Sec.~\ref{sec:current}).

\begin{figure}[t]
\includegraphics[width=\columnwidth]{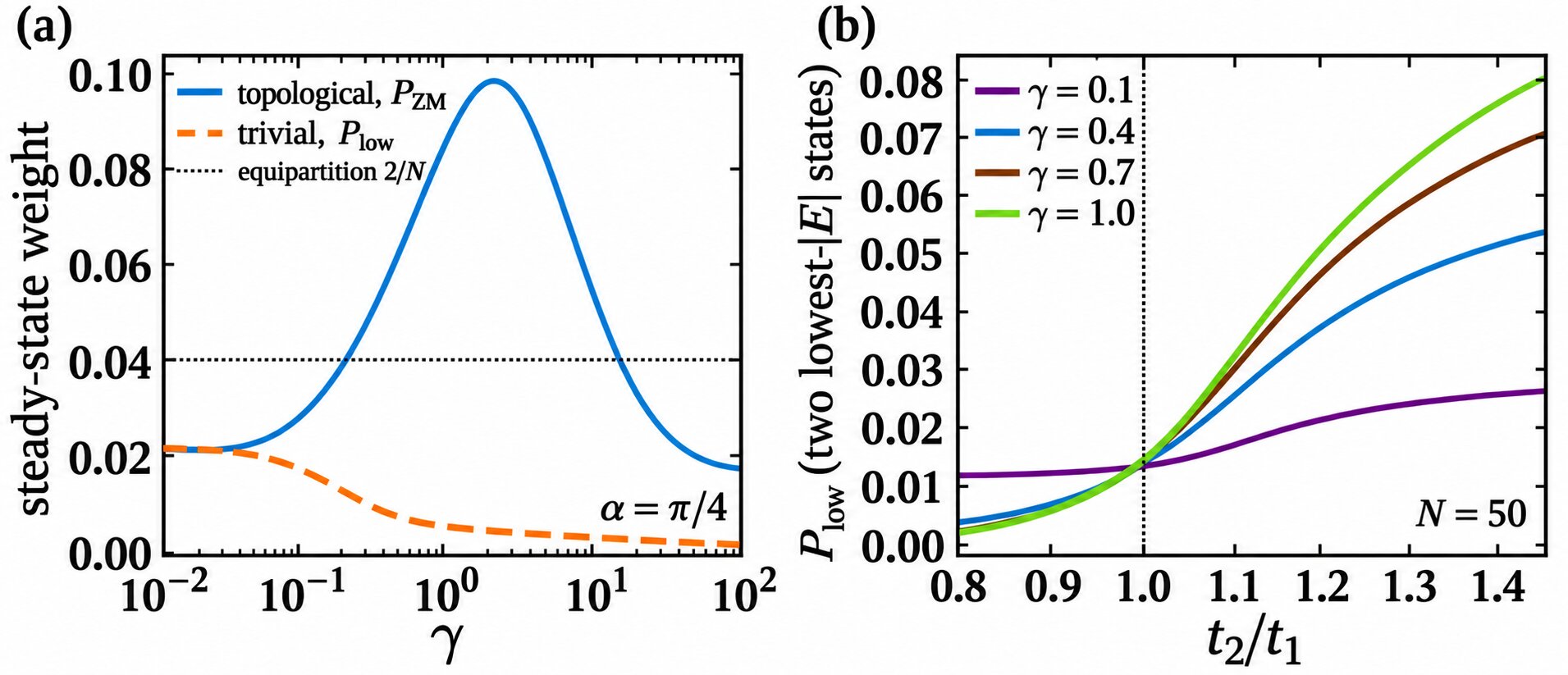}
\caption{The bath fills the zero mode at the skin edge, and the filling switches on at $t_1=t_2$. $l=1$, $\alpha=\pi/4$, $N=50$. (a)~$\PZM$ (topological, $t_1=1.5$, $t_2=2.5$), $\Plow$ (trivial, $t_1=1.5$, $t_2=1.0$), and $2/N$ versus $\gamma$. (b)~Weight of the two lowest-$|E|$ states versus $t_2/t_1$ at $t_1=1.0$.}
\label{fig:zm}
\end{figure}
\begin{figure*}[t]
\includegraphics[width=0.92\textwidth]{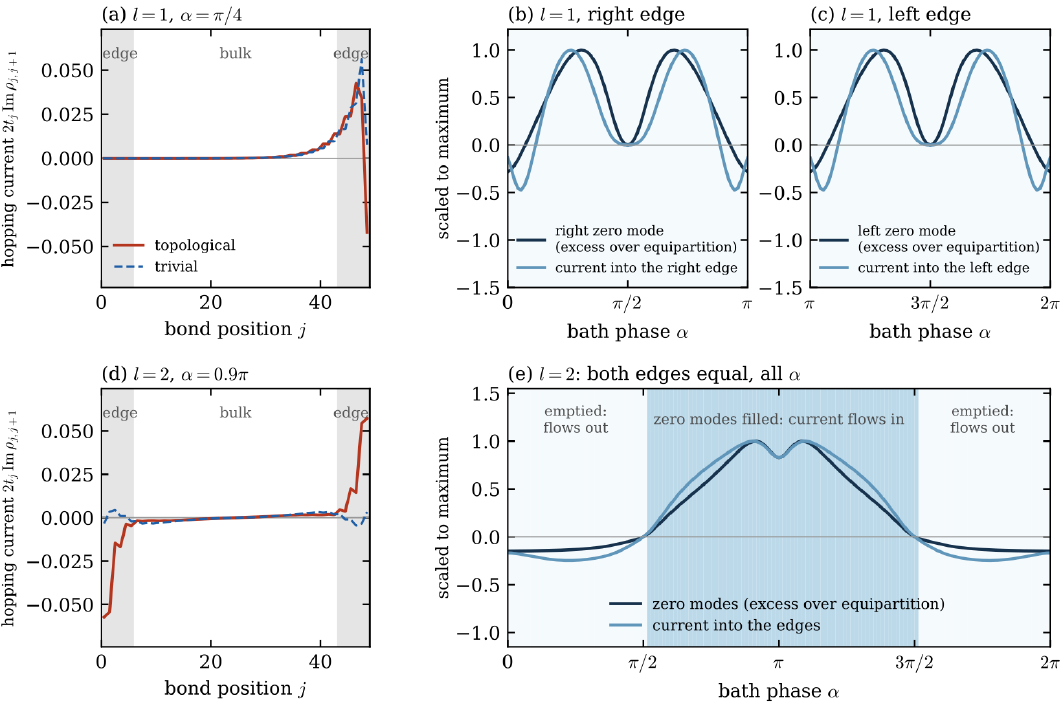}
\caption{The coherent current holds the pile-up and tracks the zero modes, as their cause for $l=1$ and as their consequence for $l=2$. (a)~Hopping current $2t_j\Imag\rho_{j,j+1}$ versus bond position for $l=1$ at $\alpha=\pi/4$, topological (solid) and trivial (dashed) phases; gray bands mark the edges. (b),(c)~Zero mode at the skin edge (excess over equipartition) and current into that edge versus $\alpha$, each scaled to its maximum (line colors as in the legends): right edge on $(0,\pi)$, left edge on $(\pi,2\pi)$. (d)~Same as (a) for $l=2$ at $\alpha=0.9\pi$: the current flows into both edges. (e)~$l=2$: zero-mode pair (excess over equipartition) and current into the edges versus $\alpha$; shading marks where the zero modes are filled and the current flows in, or emptied and it flows out. $t_1=1.0$, $t_2=1.5$ (topological) and $t_1=1.5$, $t_2=1.0$ (trivial), $\gamma=1$, $N=50$.}
\label{fig:current}
\end{figure*}

\subsection{Zero modes: skin-edge filling}
\label{sec:zm}

The weight the steady state keeps on the zero modes is $\PZM=\langle\Pi\rangle$, with $\Pi=\sum_m d_m^\dagger d_m$ the number operator built from the two eigenstates of $H$ closest to $E=0$. Since $\Pi$ commutes with $H$, the Hamiltonian cannot change this weight on its own, so $\PZM$ can only drift because of the jumps (Sec.~\ref{supp:sym}). Each jump empties one bond orbital, $v_j$, and refills another, $u_j^\dagger$. How much of this exchange reaches the zero modes is set by how much zero-mode character each orbital carries. We write $w_j=\sum_m|\{d_m,u_j^\dagger\}|^2$ for the share carried by the refilled orbital $u_j^\dagger$, and the emptied orbital $v_j$ enters through its own population $\langle v_j^\dagger v_j\rangle$. With this notation,
\begin{equation}
\begin{aligned}
\frac{d\PZM}{dt}&=\gamma\sum_j\Big[w_j\,\langle v_j^\dagger v_j\rangle-\big\langle\{v_j^\dagger v_j,\Pi\}\big\rangle\Big],\\
w_j&=\sum_m\big|\{d_m,u_j^\dagger\}\big|^2 .
\end{aligned}
\label{eq:zmbalance}
\end{equation}
The first term feeds the zero modes at a rate set by $w_j$, and the second drains them whenever the emptied orbital itself has zero-mode weight. Since the zero modes decay away from the edges, $w_j$ and the overlap with $\Pi$ are both significant only on the few bonds nearest the two ends, so $\PZM$ tracks how much steady-state weight sits on those end bonds, accurately what the skin effect controls. For nearest-neighbor jumps the zero mode at the accumulation edge should therefore gain weight, and the one at the far edge should lose it.

Figure~\ref{fig:zm}(a) confirms this behavior at $\alpha=\pi/4$. The zero-mode occupation exceeds its equipartition
value $2/N$ across nearly two decades in $\gamma$, with
weight concentrated entirely in the zero mode at the skin edge.
It remains essentially size-independent (approximately $0.09$ at $\gamma=1$ for $N=50$--$170$), while $2/N$ decreases with increasing $N$.
The maximum at intermediate $\gamma$ reflects a competition between coherent and dissipative dynamics: weak dissipation allows the
Hamiltonian to distribute the particle nearly uniformly, whereas strong dissipation drives the system toward a jump-determined state that
no longer distinguishes the edges of $H$ from the bulk. By contrast, the two lowest states of
the trivial chain, which are not edge-bound, remain below equipartition throughout.

The behavior reverses across the transition, as shown in Fig.~\ref{fig:zm}(b). For $t_2>t_1$, the projected states are localized at the two ends, allowing the edge bonds to couple strongly to them and their occupation to increase with $\gamma$. In contrast, for $t_2<t_1$, the two lowest states extend across the chain, yielding overlaps of order $1/N$ with individual bond orbitals and a progressive depletion of their occupation as $\gamma$ increases. Consequently, the curves for different $\gamma$ intersect at $t_1=t_2$, and in the thermodynamic limit,
\begin{equation}
\lim_{N\to\infty}\Plow=0\quad(t_2\le t_1),\qquad \lim_{N\to\infty}\PZM>0\quad(t_2>t_1),
\label{eq:kink}
\end{equation}
with finite-size effects rounding the resulting kink. Thus, topology determines the existence of edge states, whereas the bath controls their occupation. Remarkably, although the nearest-neighbor bath breaks the chirality protecting these states and it preferentially populates the zero mode at the skin edge while depleting its equally protected counterpart. For next-nearest-neighbor bonds, the symmetry $\mathcal{A}$ enforces equal occupation of both edges, preventing a finite share from surviving in either zero mode (Sec.~\ref{supp:sym}).

\subsection{Coherent current}
\label{sec:current}

The bath only evens out the populations of neighboring sites, so weight cannot stay gathered near one edge unless something carries it there. The only other process is coherent hopping. Under $H$, a particle moves from site $j$ to site $j+1$ with the current
\begin{equation}
J_j=2t_j\Imag\rho_{j,j+1},\qquad \rho_{j,j+1}=\langle c_{j+1}^\dagger c_j\rangle ,
\label{eq:J}
\end{equation}

which is set by the coherence $\rho_{j,j+1}$ between the two sites. It is nonzero only when the amplitudes on the two sites differ in phase, and the bath phase $\alpha$ enters precisely through such phase differences between the sites of a bond. In an open-chain steady state no weight leaves through the ends, so the current must cancel the evening-out by the bath,
\begin{equation}
J_j=\begin{cases}
\gamma\,(n_{j+1}-n_j), & l=1,\\
\gamma\,(n_{j+1}+n_{j+2}-n_{j-1}-n_j), & l=2,
\end{cases}
\label{eq:balance}
\end{equation}
with $n_j$ the steady-state population of site $j$ and terms on nonexistent sites omitted (Sec.~\ref{supp:sym}). The bath moves weight from the more crowded site to the less crowded one, and the current carries it back.

Equation~\eqref{eq:balance} links the coherences to the populations, and through them to the zero modes. For $l=1$, adding it up from the bulk to an end shows that the extra population at the end equals the current collected along the way divided by $\gamma$. The zero modes of $H$ sit on the last few sites and are fed by the population there (Sec.~\ref{sec:zm}). The zero mode at an edge should therefore be filled by the current flowing into that edge, and only when the steady state carries coherence. The zero modes cannot carry this current themselves, since each lives on a single sublattice and has no coherence between neighboring sites.

\emph{Current as the origin of nearest-neighbor jumps accumulation.-} On a ring, the steady-state current is uniform across all bonds. Since the bath-induced contributions $\gamma(n_j-n_{j+1})$ cancel within each unit cell, the net current,
\begin{equation}
J=\tfrac12\big(2t_1\Imag\rho_{AB}+2t_2\Imag\rho_{BA'}\big),
\label{eq:Jring}
\end{equation}
arises entirely from coherent hopping, mediated by bond coherences imprinted by the bath phase $\alpha$, where $\rho_{AB}$ and $\rho_{BA'}$ denote the intracell and intercell coherences, respectively. Both mirror reflection, which maps $\alpha\to-\alpha$, and the symmetry $\mathcal{A}$, which shifts $\alpha\to\alpha+\pi$, reverse the current. Consequently, $J$ vanishes at $\alpha=0,\pi$, where inversion symmetry is restored, and at the Hermitian points, where the steady state reduces to the identity and all coherences vanish. Upon disconnecting the ring, the circulating current is interrupted and particles accumulate toward the edge favored by the current. Consistently, the current direction matches the open-chain accumulation direction for all tested values of $\alpha$, irrespective of the bare Hamiltonian's topological phase.

On the open chain the current is appreciable only near the edge, where the populations change from site to site [Fig.~\ref{fig:current}(a)]. Away from the last bond or two it is nearly the same in the topological and trivial chains, as expected since the winding $W_0$ belongs to the bath. The same current fills the zero mode at the skin edge. The population that the current gathers at the end is what the end bonds draw on to feed the zero mode, so as $\alpha$ is varied the zero-mode weight above equipartition should follow the current into that edge. Figures~\ref{fig:current}(b) and \ref{fig:current}(c) confirm this. The two curves change sign at nearly the same phases, peak at similar phases, and vanish at the Hermitian point, where the steady state is the identity and carries no coherence. They are similar in shape but not identical. Panel (c) is panel (b) shifted by $\pi$, since $\mathcal{A}$ relates $\alpha$ and $\alpha+\pi$ and thereby exchanges the two edges. The filling of the zero mode is thus a consequence of the current driven by the bath phase, and not a property of the zero modes alone.
\begin{figure}[t]
\includegraphics[width=\columnwidth]{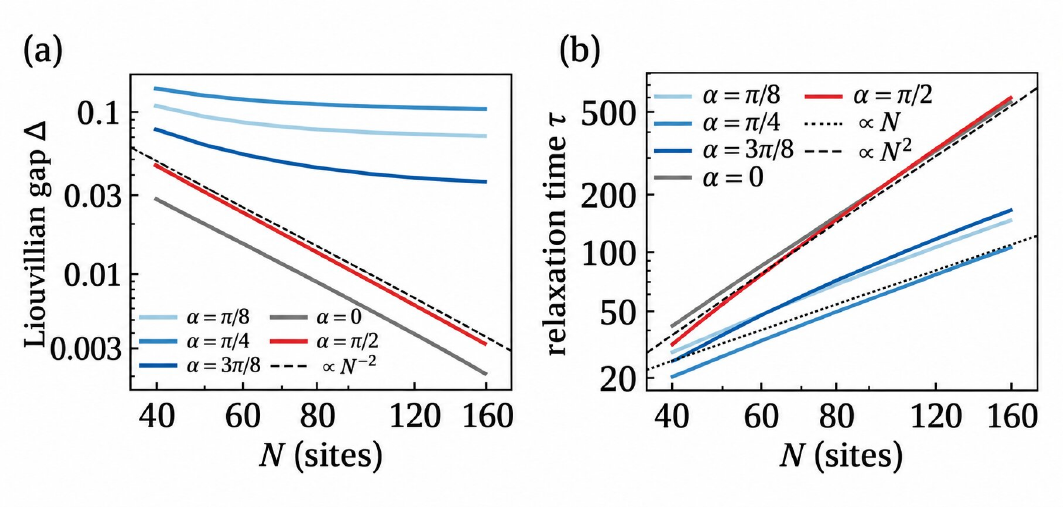}
\caption{Relaxation is linear in size only in the skin regime. $l=1$, $t_1=1.0$, $t_2=1.5$, $\gamma=1$. (a)~Gap $\Delta$ and (b)~relaxation time $\tau$ versus $N$; black lines indicate $N^{-2}$, $N$, and $N^{2}$.}
\label{fig:relaxN}
\end{figure}

\emph{Current as a consequence for next-nearest neighbor jumps ---}For $l=2$ the symmetry $\mathcal{A}$ holds at every $\alpha$, so the two ends of the chain are equivalent. A current that flows toward the right at the right end must flow toward the left at the left end. It therefore runs into both ends or away from both, and no current can circulate around a ring [Fig.~\ref{fig:current}(d)]. Here the zero modes are not filled by a current. They are filled directly by the energy selection of Sec.~\ref{sec:energy}, which favors their sign pattern around $\alpha=\pi$. Once filled, they leave the last sites more populated than their neighbors, the bath tries to even this out, and a coherent current from the bulk has to run into the ends to keep it in place. The current into the edges then follows the zero-mode excess [Fig.~\ref{fig:current}(e)]. It flows inward for $\pi/2<\alpha<3\pi/2$, where the zero modes hold more than equipartition, and outward elsewhere, equally at both ends. The trivial chain has no zero modes and carries almost no current at its edges.
\section{Relaxation dynamics}
\label{sec:relax}
We finally ask how the density skin effect shows up in the approach to the steady state, restricting the discussion to $l=1$. Following Haga \emph{et al.}~\cite{Haga2021}, we start a particle at the left end, $\rho(0)=c_0^\dagger\vac\langle\mathrm{vac}|c_0$, monitor the density at the right end, and define $\tau$ as the latest time at which its deviation from the steady value exceeds $e^{-1}$ of the initial deviation. In the skin regime the steady state sits at the right end, so this protocol starts the particle as far as possible from where it ends up. The asymptotic decay is set by the Liouvillian gap $\Delta=|\Real\lambda_1|$, but the gap alone does not fix $\tau$. If the slowest right and left eigenmodes are localized at opposite edges with length $\xi$, their overlap is exponentially small, a left-edge initial state needs an exponentially large amplitude in the slowest mode, and this amplitude must decay at rate $\Delta$~\cite{Haga2021},
\begin{equation}
\tau\sim\Delta^{-1}\big(1+N/\xi\big).
\label{eq:haga}
\end{equation}
Here $\xi$ is the localization length of the steady state, read off from the exponential decay of the density away from the skin edge. The argument uses only the fact that the steady state and the slowest modes are localized at an edge, not how that localization arises. It therefore applies to our bath, whose transfer amplitudes have equal magnitude in both directions, exactly as it does to the asymmetric hopping of Ref.~\cite{Haga2021}: once the interference of Eq.~\eqref{eq:interference} has produced a skin effect, the relaxation follows the same law.

In the skin regime the gap stays finite, $\Delta\to\Delta_\infty(\alpha)$, and $\xi(\alpha)$ is finite and independent of $N$ (about $3.8$ sites at $\alpha=\pi/4$), so Eq.~\eqref{eq:haga} predicts relaxation linear in size,
\begin{equation}
\tau\simeq\frac{N}{\xi(\alpha)\,\Delta_\infty(\alpha)} .
\label{eq:tauskin}
\end{equation}


\begin{figure}[t]
\includegraphics[width=\columnwidth]{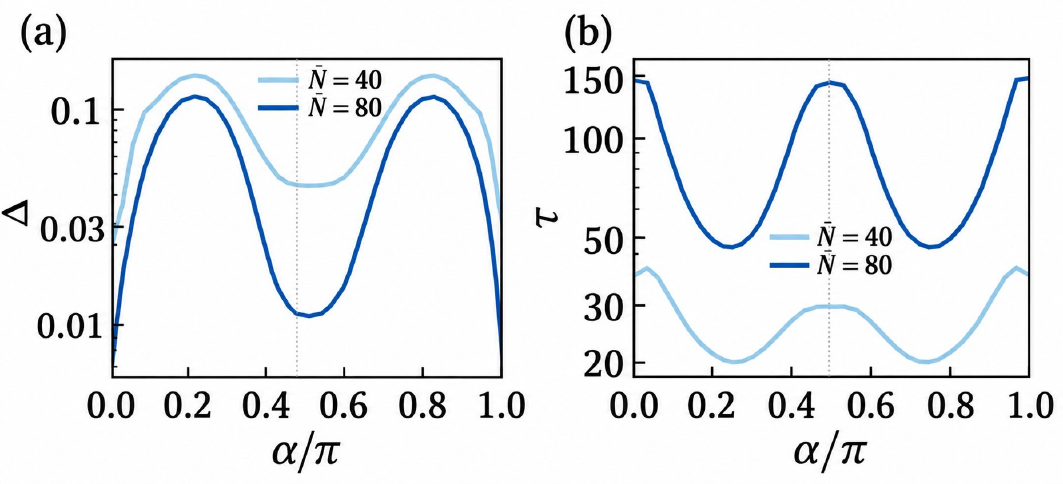}
\caption{The relaxation is symmetric about $\alpha=\pi/2$ and fastest near $\alpha=\pi/4$ and $3\pi/4$. (a)~$\Delta$ and (b)~$\tau$ versus $\alpha$ for $N=40$ and $80$; other parameters as in Fig.~\ref{fig:relaxN}.}
\label{fig:relaxA}
\end{figure}

Figs.~\ref{fig:relaxN}(a) and \ref{fig:relaxN}(b) confirm this: $\Delta$ saturates, $\tau$ grows linearly, and $\tau\Delta_\infty\xi/N\approx0.2$ for all skin-regime values of $\alpha$ shown ($\xi$ measured in sites). At $\alpha=0$ and $\pi$ there is no edge accumulation, the gap closes as $\Delta\propto N^{-2}$ and $\tau\simeq\Delta^{-1}\propto N^2$, which is Eq.~\eqref{eq:haga} with $N/\xi\to0$ (diffusive spreading). The dependence on $\alpha$ [Figs.~\ref{fig:relaxA}(a) and \ref{fig:relaxA}(b)] is symmetric under $\alpha\to\pi-\alpha$, which follows from combining $\mathcal{A}$ with the mirror. Relaxation is fastest near $\alpha=\pi/4$, where the relaxation velocity $N/\tau\sim\xi\Delta_\infty$ is largest. Because $\tau$ is linear in $N$ there and quadratic at $\alpha=\pi/2$, the Hermitian point is already about three times slower than $\alpha=\pi/4$ at $N=80$, and the ratio keeps growing with $N$.

\emph{Hermitian points.---}The points $\alpha=\pi/2+n\pi$ are an exception to
Eq.~\eqref{eq:tauskin}. Here the jumps are Hermitian, $O_j^\dagger=O_j$, and the
open-chain eigenmodes are still edge localized, with skin character. However, they occur
in mirror pairs: the adjoint Liouvillian, which propagates the left eigenmodes, is the
mirror image of the complex-conjugated Liouvillian (Sec.~\ref{supp:sym}),
\begin{equation}
\Lv^\dagger(X)=P\big[\Lv(PX^{*}P)\big]^{*}P\qquad(\alpha=\pi/2+n\pi),
\label{eq:mirrorDB}
\end{equation}
so the left eigenmode belonging to each right eigenmode sits at the opposite edge, as in any skin effect. The right eigenmodes themselves stay localized at the same edge as for neighboring $\alpha$. What the Hermitian point removes is the localization of the steady state, not that of the decaying modes.

A Hermitian bath has a more basic consequence. Hermitian jumps always leave the
identity invariant, $\Lv(\id)=0$, for any $H$, so the steady state is the featureless
mixture $\rss=\id/N$ [Eq.~\eqref{eq:unital}]. The argument of Haga \emph{et al.}\
requires the steady state and the slowest decaying mode to sit at an edge: only then
does an initial state at the opposite edge need an exponentially large amplitude in the
slowest mode. At the Hermitian points both are extended, so $N/\xi\to0$ in
Eq.~\eqref{eq:haga} and the relaxation is set by the gap alone, $\tau\simeq\Delta^{-1}$.
Because $\Delta\propto N^{-2}$, the relaxation is diffusive, $\tau\propto N^{2}$
[Fig.~\ref{fig:relaxN}]. Edge-localized eigenmodes therefore do not by themselves
produce skin-type relaxation; what matters is whether the steady state is localized,
and a Hermitian bath always leaves it unital.

\begin{figure}[t]
\includegraphics[width=\columnwidth]{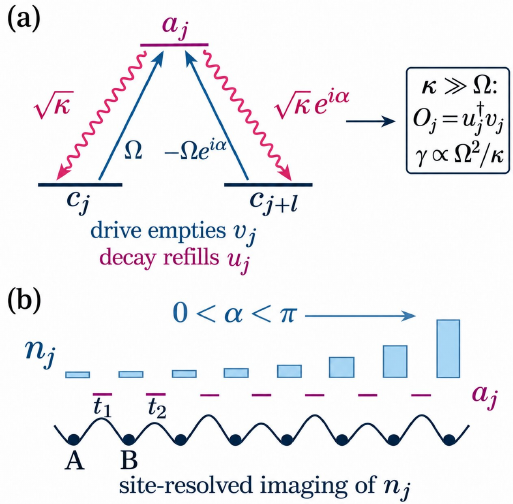}
\caption{Proposed realization. (a)~A drive couples the bond mode $v_j$ (amplitudes $\Omega$ and $-\Omega e^{i\alpha}$ on the sites $j$ and $j+l$) to a short-lived auxiliary mode $a_j$, which decays at rate $\kappa$ into $u_j$ (amplitudes $1$ and $e^{i\alpha}$); for $\kappa\gg\Omega$ this realizes $O_j$ with $\gamma\propto\Omega^2/\kappa$. (b)~Dimerized optical lattice with an auxiliary level for every bond ($l=1$); the steady-state density $n_j$ piles up at the right edge for $0<\alpha<\pi$.}
\label{fig:experiment}
\end{figure}
\section{Proposed experimental realization}
\label{sec:experiment}

In order to connect the theory with a concrete setting, we outline how the bond jump~\eqref{eq:O} can be engineered and how its consequences can be detected. From a microscopic point of view, the jump factorizes into two steps [Fig.~\ref{fig:experiment}(a)]: a coherent drive that transfers the particle from the bond mode $v_j$ to a short-lived auxiliary mode $a_j$, and a decay of $a_j$ that returns it into $u_j$,
\begin{equation}
H_d=\Omega\sum_j\big(a_j^\dagger v_j+\mathrm{h.c.}\big),\qquad L_j=\sqrt{\kappa}\,u_j^\dagger a_j,
\label{eq:scheme}
\end{equation}
with $v_j$ and $u_j^\dagger$ of Eq.~\eqref{eq:rank1}. For $\kappa\gg\Omega$, adiabatic elimination of the auxiliary modes~\cite{Reiter2012} yields Eq.~\eqref{eq:O} with $\gamma\propto\Omega^2/\kappa$. This is the structure of reservoir-engineering schemes for cold atoms~\cite{Griessner2006,Diehl2008,Kraus2008}: the auxiliary mode is an excited level held in a state-dependent lattice shifted by half a period, the drive is a laser whose phase pattern sets the relative phase in $v_j$, and the decay proceeds by spontaneous emission or by emission into a reservoir gas. Explicit cold-atom schemes for these bond jumps with $l=1$ and $l=2$ have been proposed in Ref.~\cite{Liu2025}, and related proposals exist for superconducting resonator arrays~\cite{Marcos2012}.

In such schemes, isotropic emission refills the in-phase combination, which corresponds to $e^{i\alpha}=1$ in $u_j^\dagger$. A general phase $\alpha$ is nevertheless accessible. The single-site gauge transformation $c_j\to e^{i\alpha j/l}c_j$ maps the jump~\eqref{eq:O} onto
\begin{equation}
O_j\ \mapsto\ \big(c_j^\dagger+c_{j+l}^\dagger\big)\big(c_j-e^{2i\alpha}c_{j+l}\big),
\label{eq:gauge}
\end{equation}
at the expense of a uniform Peierls phase $\alpha/l$ per lattice spacing in $H$. Since this transformation is diagonal in the site basis, the open-chain densities are unchanged. An in-phase refill, a drive with relative phase $\pi+2\alpha$ between the bond sites, and a synthetic gauge field realized by laser-assisted tunneling~\cite{Aidelsburger2013,Miyake2013} therefore reproduce the model at arbitrary $\alpha$. Dimerized optical lattices of the required form have been realized and characterized topologically~\cite{Atala2013}.

The steady-state signatures discussed above are accessible with site-resolved imaging in quantum gas microscopes~\cite{Bakr2009,Sherson2010}: the density profile reveals the pile-up and its reversal under $\alpha\to\alpha+\pi$, and the coherent current can be measured with the double-well projection technique of Ref.~\cite{Atala2014}. Since the dynamics is restricted to a single particle, the particle statistics play no role, and arrays of qubits with a single excitation offer an alternative platform~\cite{Barreiro2011,Harrington2022,Mi2024}. There, the vacuum-sector coherence of Sec.~\ref{sec:twoindex} becomes a local observable, $\rho_{i,\mathrm{vac}}=\langle\sigma_i^-\rangle$, so that the opposite windings of the two vacuum loops for $l=2$ can be probed by single-qubit tomography.

\section{Summary and outlook}
\label{sec:summary}
We have shown that the number-conserving jump operators with a  tunable phase generate a Liouvillian skin effect in dimerized chain, even if their exchange of population is balanced. The push originates from the interference between the on-site and transfer parts of each jump operators and is fixed by the phase $\alpha$ through the spectral winding $W_0(\alpha)$, which we evaluated from the closed-form $K$-sector Liouvillian $\Lv(K)$ for the nearest-neighbor bonds and which belongs to the bath rather than to the band topology of the bare Hamiltonian. Resolving the winding into the two indices of the density matrix revealed two distinct skin effects, decided by a single antiunitary mirror-chiral symmetry. For $l=1$ the jumps operators break it, the ket and bra are pushed together, producing a skin effect of the density. For $l=2$ the jumps preserve however it, the ket and bra are pinned to opposite edges, producing a bipolar skin effect of coherences that resides in the fastest relaxing eigen modes. The density skin effect is sustained by a coherent current, fills the zero mode at the skin edge with a size-independent weight that switches on at $t_1=t_2$, and makes relaxation linear in system size except at the Hermitian points.

Several questions remain open like an analytic expression for $W_0(\alpha)$ beyond its symmetry constraints, Eq.~\eqref{eq:W0sym}, is lacking. The necessary ingredients identified here, interference, coherent hopping, broken inversion, and non-unitality, may serve as a starting point for a general criterion for skin effects induced by balanced jumps~\cite{Shimomura2024}. The relation between the pair $(\nK,\nB)$ and the localization corner of each eigenmode is supported numerically but not yet proven. Finally, the jumps are quadratic in the fermion operators, so that the Lindbladian is not quadratic and the many-particle problem lies beyond the Gaussian frameworks of Refs.~\cite{Prosen2008,Lieu2020} and the fate of the two skin effects at finite filling remains to be explored.


\bibliography{references}

\clearpage
\onecolumngrid
\begin{center}
{\large\bfseries Supplementary Material}
\end{center}
\setcounter{section}{0}\renewcommand{\thesection}{S\arabic{section}}
\setcounter{figure}{0}\renewcommand{\thefigure}{S\arabic{figure}}
\setcounter{table}{0}\renewcommand{\thetable}{S\arabic{table}}
\setcounter{equation}{0}\renewcommand{\theequation}{S\arabic{equation}}


\makeatletter
\def\subsection{\@startsection{subsection}{2}{\z@}{.9cm \@plus.6ex \@minus .2ex}{.4cm}{\normalfont\small\bfseries\centering}}
\makeatother
\setlength{\parskip}{0.5ex plus 0.3ex}
\setlength{\abovedisplayskip}{9pt plus 3pt minus 3pt}
\setlength{\belowdisplayskip}{9pt plus 3pt minus 3pt}
\renewcommand{\arraystretch}{1.15}
\setlength{\textfloatsep}{16pt plus 4pt minus 4pt}

\begin{center}
\begin{minipage}{0.82\textwidth}
\small
\noindent\textbf{Contents}\par\smallskip
\noindent S1.\ Liouvillian windings: the $K$ sector and ket and bra twists \dotfill \pageref{supp:LKk}\\
\hspace*{1.5em}A.\ The $K$ sector for $l=1$ \dotfill \pageref{supp:Ksector}\\
\hspace*{1.5em}B.\ Closed form, determinant, and $W_0$ \dotfill \pageref{supp:W0}\\
\hspace*{1.5em}C.\ Ket and bra twists \dotfill \pageref{supp:twist}\\[2pt]
S2.\ Symmetry and balance \dotfill \pageref{supp:sym}\\
\hspace*{1.5em}A.\ Single-jump identities and balance relations \dotfill \pageref{supp:balance}\\
\hspace*{1.5em}B.\ Windings under the mirror, Hermiticity, and $\mathcal{A}$ \dotfill \pageref{supp:maps}\\
\hspace*{1.5em}C.\ Steady state, Hermitian points, and zero modes \dotfill \pageref{supp:ss}\\[2pt]
S3.\ Vacuum sector, corner rule, and skin modes \dotfill \pageref{supp:vacuum}\\
\hspace*{1.5em}A.\ Blocks and vacuum loops \dotfill \pageref{supp:loops}\\
\hspace*{1.5em}B.\ Ket-winding shading and the transient coherence \dotfill \pageref{supp:corner}
\end{minipage}
\end{center}
\bigskip

\section{Liouvillian windings: the \texorpdfstring{$K$}{K} sector and ket and bra twists}
\label{supp:LKk}

We first construct the Liouvillian of the periodic $l=1$ chain in the sector of fixed quasimomentum transfer, following Ref.~\cite{Long2026}. We then define the twisted ring of Ref.~\cite{Hamanaka2023} separately for the ket and the bra index, which applies to both ranges, and show that the two constructions agree at the steady-state reference.

\subsection{The \texorpdfstring{$K$}{K} sector for \texorpdfstring{$l=1$}{l=1}}
\label{supp:Ksector}

\emph{Conventions.} The ring has $L$ unit cells $m=1,\dots,L$ with sites $(A_m,B_m)$. The two $l=1$ jumps of cell $m$ are
\begin{equation}
O_m^{\mathrm{intra}}=\big(c_{A,m}^\dagger+z\,c_{B,m}^\dagger\big)\big(c_{A,m}-z\,c_{B,m}\big),\qquad
O_m^{\mathrm{inter}}=\big(c_{B,m}^\dagger+z\,c_{A,m+1}^\dagger\big)\big(c_{B,m}-z\,c_{A,m+1}\big).
\end{equation}
With $c_{\sigma,m}=L^{-1/2}\sum_k e^{ikm}c_{\sigma,k}$ ($\sigma=A,B$) and $k_n=2\pi n/L$, the Bloch Hamiltonian is
\begin{equation}
H(k)=\begin{pmatrix}0 & h(k)\\ h^*(k) & 0\end{pmatrix},\qquad h(k)=t_1+t_2e^{-ik}.
\end{equation}
We collect the coherences $\rho_{\sigma\sigma'}(k,k')=\langle c_{\sigma',k'}^\dagger c_{\sigma,k}\rangle$ into $2\times2$ sublattice blocks $\rho(k,k')$, vectorized in the order $(AA,AB,BA,BB)$ with $\mathrm{vec}(X\rho Y)=(X\otimes Y^{T})\,\mathrm{vec}\,\rho$. A translation by one cell multiplies $\rho(k,k')$ by $e^{i(k'-k)}$ and leaves $\Lv$ invariant, so $\Lv$ is block diagonal in $K=k'-k$. Within a sector, blocks are labeled by the internal momentum $k$ (with $k'=k+K$). We write $\Lv(K,k)$ for the part of $\Lv$ that acts within one block, while the jump terms couple different blocks of the same sector. Operators such as $O=\sum_{\sigma\sigma'}c_\sigma^\dagger(O)_{\sigma\sigma'}c_{\sigma'}$ are represented by their coefficient matrices.

\emph{Hamiltonian and loss terms.} The Hamiltonian part of the block $(k,k')$ is
\begin{equation}
\Lv_H(k,k')=-i\big[H(k)\otimes\id-\id\otimes H^{T}(k')\big]
=\begin{pmatrix}
0 & ih^*(k') & -ih(k) & 0\\
ih(k') & 0 & 0 & -ih(k)\\
-ih^*(k) & 0 & 0 & ih^*(k')\\
0 & -ih^*(k) & ih(k') & 0
\end{pmatrix}.
\label{eq:LH}
\end{equation}
The intracell jump carries no momentum dependence, and its coefficient matrix and that of $O^\dagger O$ are
\begin{equation}
O^{\mathrm{intra}}=\begin{pmatrix}1 & -z\\ z & -z^2\end{pmatrix},\qquad
O^{\mathrm{intra}\dagger}O^{\mathrm{intra}}=2\begin{pmatrix}1 & -z\\ -z^* & 1\end{pmatrix},
\end{equation}
and its loss term is $-\tfrac{\gamma}{2}\big[O^\dagger O\otimes\id+\id\otimes(O^\dagger O)^{T}\big]$. The intercell jump at equal input and output momentum is
\begin{equation}
O_I(k)=\begin{pmatrix}-z^2 & z\,e^{-ik}\\ -z\,e^{ik} & 1\end{pmatrix},\qquad
O_I^\dagger O_I(k)=2\begin{pmatrix}1 & -z^*e^{-ik}\\ -z\,e^{ik} & 1\end{pmatrix},
\end{equation}
with loss term $-\tfrac{\gamma}{2}\big[O_I^\dagger O_I(k)\otimes\id+\id\otimes\big(O_I^\dagger O_I(k')\big)^{T}\big]$. The diagonal block is
\begin{equation}
\Lv(K,k)=\Lv_H(k,k+K)+\Lv^{\mathrm{intra}}_{\mathrm{loss}}+\Lv^{\mathrm{inter}}_{\mathrm{loss}}(k,k+K).
\end{equation}

\emph{Jump (recycling) terms.} Because $O^{\mathrm{intra}}$ is local and momentum independent, $\sum_m O_m\rho O_m^\dagger$ couples every pair of blocks of a sector with the same matrix,
\begin{equation}
M_0^{\mathrm{intra}}=\frac{\gamma}{L}\,O\otimes O^*=\frac{\gamma}{L}
\begin{pmatrix}
1 & -z^* & -z & 1\\
z^* & -(z^*)^2 & -1 & z^*\\
z & -1 & -z^2 & z\\
1 & -z^* & -z & 1
\end{pmatrix}.
\label{eq:M0intra}
\end{equation}
A jump removes the particle from one Bloch pair and refills a local bond orbital, which contains every momentum. For the intercell bond, $O_m^{\mathrm{inter}}=c_{B,m}^\dagger c_{B,m}-z\,c_{B,m}^\dagger c_{A,m+1}+z\,c_{A,m+1}^\dagger c_{B,m}-z^2c_{A,m+1}^\dagger c_{A,m+1}$ in real space, and in momentum space
\begin{equation}
O_m^{\mathrm{inter}}=\frac1L\sum_{k,q}e^{-i(k-q)m}\,c_k^\dagger O_I(k,q)\,c_q,\qquad
O_I(k,q)=\begin{pmatrix}-z^2e^{-i(k-q)} & z\,e^{-ik}\\ -z\,e^{iq} & 1\end{pmatrix},
\end{equation}
with $O_I(k,k)\equiv O_I(k)$. The sum over $m$ conserves only $K$, so the block $(k,k+K)$ receives a contribution from every block $(q,q+K)$ of the sector,
\begin{equation}
\frac{\gamma}{L}\,O_I(k,q)\otimes O_I^*(k+K,q+K)=S_k\,\tilde M_0(K)\,S_q^{-1},
\label{eq:Mtilde_def}
\end{equation}
with gauge factors $S_k=\mathrm{diag}(1,e^{-ik},e^{ik},1)$ and
\begin{equation}
\tilde M_0(K)=\frac{\gamma}{L}
\begin{pmatrix}
1 & -z\,e^{iK} & -z^* & e^{iK}\\
z\,e^{-iK} & -z^2 & -e^{-iK} & z\\
z^* & -e^{iK} & -(z^*)^2 & z^*e^{iK}\\
e^{-iK} & -z & -z^*e^{-iK} & 1
\end{pmatrix}.
\label{eq:Mtilde}
\end{equation}

\subsection{Closed form, determinant, and \texorpdfstring{$W_0$}{W0}}
\label{supp:W0}

Collecting the pieces gives Eq.~\eqref{eq:LK} of the main text, with $S=\mathrm{diag}[S_{k_1},\dots,S_{k_L}]$; the $k=q$ parts of the jump terms sit on the diagonal of $N_L$. To evaluate the determinant we write $\Lv(K)-\lambda=A+UV^{T}$ with
\begin{equation}
A=\textstyle\bigoplus_n\big[\Lv(K,k_n)-\lambda\big],\qquad
U=\big[\mathbf{1}_L\otimes I_4,\ S(\mathbf{1}_L\otimes I_4)\big],\qquad
V^{T}=\begin{bmatrix}\mathbf{1}_L^{T}\otimes M_0^{\mathrm{intra}}\\ (\mathbf{1}_L^{T}\otimes\tilde M_0)S^{-1}\end{bmatrix},
\end{equation}
where $\mathbf{1}_L$ is the all-ones vector. With $R_n=[\Lv(K,k_n)-\lambda]^{-1}$ the matrix determinant lemma gives
\begin{equation}
D(K,\lambda)=\prod_n\det\big[\Lv(K,k_n)-\lambda\big]\,\det\big[I_8+Q(K,\lambda)\big],\qquad
Q=\sum_n\begin{pmatrix}M_0^{\mathrm{intra}}R_n & M_0^{\mathrm{intra}}R_nS_{k_n}\\ \tilde M_0S_{k_n}^{-1}R_n & \tilde M_0S_{k_n}^{-1}R_nS_{k_n}\end{pmatrix},
\label{eq:Dlemma}
\end{equation}
which reduces the determinant of the $4L\times4L$ sector matrix to $L$ determinants of size $4$ and a single one of size $8$.

\emph{The winding.} On a discrete grid, Eq.~\eqref{eq:W0} reads
\begin{equation}
W_0=\frac{1}{2\pi}\sum_m\mathrm{Arg}\,\frac{D(K_{m+1},0^-)}{D(K_m,0^-)},
\end{equation}
with $0^-$ a reference point just below zero. Close to the Hermitian points the steady-state branch passes this reference tangentially, so the grid must resolve the neighborhood of $K=0$. The result is Eq.~\eqref{eq:W0piece} of the main text, with $W_0>0$ corresponding to the right edge. At $\alpha=\pi/2+n\pi$ the evaluated value is reversed [Fig.~\ref{fig:skin}(b)]. Since the spectrum touches $0^-$ at these points and the value depends on the reference offset, it cannot be regarded as an invariant.

\emph{Symmetry constraints in $K$ space.} The mirror reverses both momenta, so it maps $\Lv_\alpha(K)$ unitarily onto $\Lv_{-\alpha}(-K)$. The symmetry $\mathcal{A}$ of Eq.~\eqref{eq:sym} combines the mirror with complex conjugation, which also reverses both momenta, and with the sublattice sign, which leaves them unchanged; it maps $\Lv_\alpha(K)$ antiunitarily onto $\Lv_{\alpha+\pi}(K)$. Both maps preserve the $k$ grid. For real $E$,
\begin{equation}
D_{-\alpha}(-K,E)=D_\alpha(K,E),\qquad D_{\alpha+\pi}(K,E)=D_\alpha(K,E)^*.
\end{equation}
Running $K$ backward reverses the winding, and so does complex conjugation. Hence $W_0(-\alpha)=-W_0(\alpha)$ and $W_0(\alpha+\pi)=-W_0(\alpha)$, Eq.~\eqref{eq:W0sym}; with $2\pi$ periodicity, $W_0(0)=W_0(\pi)=0$.

\subsection{Ket and bra twists}
\label{supp:twist}

Following Ref.~\cite{Hamanaka2023}, we close the chain into a ring of $N$ sites and twist the boundary. Let $p=N-1$ and $q=0$ be the sites on either side of the boundary bond. In every operator $X\in\{H,O_j,O_j^\dagger O_j\}$, the terms that move a particle forward across the boundary, $c_q^\dagger c_p$, are multiplied by $e^{i\theta}$, and their partners $c_p^\dagger c_q$ by $e^{-i\theta}$; we denote the result by $X_\theta$. For $l=2$ the bath bonds $(N-2,0)$ and $(N-1,1)$ also cross the boundary and are twisted in the same way. In the Lindblad equation~\eqref{eq:lindblad} the operators to the left of $\rho$ act on the ket and those to the right on the bra (Table~\ref{tab:twist}); twisting them separately gives
\begin{equation}
\Lv(\theta_{\mathrm K},\theta_{\mathrm B})\rho=-i\big(H_{\theta_{\mathrm K}}\rho-\rho H_{-\theta_{\mathrm B}}\big)
+\gamma\sum_j O_{j,\theta_{\mathrm K}}\,\rho\,\big(O_{j,-\theta_{\mathrm B}}\big)^\dagger
-\frac{\gamma}{2}\sum_j\Big[\big(O_j^\dagger O_j\big)_{\theta_{\mathrm K}}\rho+\rho\big(O_j^\dagger O_j\big)_{-\theta_{\mathrm B}}\Big].
\label{eq:Ltwist}
\end{equation}
The ket winding $\nK$ uses the loop $(\theta,0)$, the bra winding $\nB$ the loop $(0,\theta)$, and the density winding $\nC$ the loop $(\theta,\theta)$, all for $\theta$ from $0$ to $2\pi$, as in Eq.~\eqref{eq:nu}. On the bra side the twist carries $-\theta_{\mathrm B}$, because in $\rho X$ the bra index is the first index of $X$, so that the term moving the bra index forward across the boundary is $c_p^\dagger c_q$, the transpose of the ket term. With this convention $\nK$ and $\nB$ both count displacement to the right; placing $+\theta_{\mathrm B}$ instead would reverse every bra winding and turn Eq.~\eqref{eq:Crule} into $\nB=+\nK$. At $\theta_{\mathrm K}=\theta_{\mathrm B}=0$, Eq.~\eqref{eq:Ltwist} is the ordinary ring Liouvillian.

\begin{table}[t]
\caption{Terms that carry the twist phases ($p=N-1$, $q=0$; $X=H$, $O_j$, or $O_j^\dagger O_j$). Hermitian partners carry the conjugate phase.}
\label{tab:twist}
\begin{ruledtabular}
\begin{tabular}{lll}
Term & Index that moves & Phase\\
\hline
$H\rho$, $O\rho O^\dagger$, $O^\dagger O\rho$ & ket, $p\to q$ & $c_q^\dagger c_p$ in $X$: $\times e^{i\theta_{\mathrm K}}$\\
$\rho H$, $\rho\, O^\dagger O$ & bra, $p\to q$ & $c_p^\dagger c_q$ in $X$: $\times e^{i\theta_{\mathrm B}}$\\
$O\rho O^\dagger$ & bra, $p\to q$ & $c_p^\dagger c_q$ in $O^\dagger$: $\times e^{i\theta_{\mathrm B}}$
\end{tabular}
\end{ruledtabular}
\end{table}

\emph{Counting.} A winding counts eigenvalue paths encircling $\Lambda$ counterclockwise minus clockwise. With the phase on the forward hop, $\nu>0$ means accumulation at the right edge. Only signs are quoted, because magnitudes grow with the ring size. The ket-only twist reproduces the winding of Ref.~\cite{Hamanaka2023}.

\emph{Relation of the twists to the $K$ sector.}\label{supp:twistK} The relation $W_0=\nK(0^-)=\nB(0^-)=\nC(0^-)/2$, Eq.~\eqref{eq:nuW0}, follows in five steps.
(1)~A phase $e^{i\theta}$ on the boundary ket hops is equivalent to $e^{i\theta/N}$ on every ket hop, since the transformation $c_j\to e^{i\theta j/N}c_j$ applied to the ket side alone is a similarity transformation of $\Lv$ and does not change $\det[\Lv(\theta)-\Lambda]$.
(2)~A phase on every ket hop shifts every ket momentum by $\delta=\theta/L$ while the bra momenta stay put, so that each pair $(k,k')$ moves to $K+\delta$ and the whole sector slides along the $K$ axis.
(3)~At $\theta=2\pi$ one has $\delta=2\pi/L$, so that every ket momentum lands on the next grid point and every sector $K_m$ has moved to its neighbor. The spectrum is thus mapped onto itself with the sectors relabeled.
(4)~$\det[\Lv(\theta)-\Lambda]$ is the product of the sector determinants, so its total phase change is the sum over $m$ of the change along $K_m\to K_{m\pm1}$, which summed over all $m$ is one trip around the $K$ circle, i.e., the sum defining $W_0$. Hence $\nK(0^-)=W_0$. The twist supplies the continuous path between grid points that the discrete sum samples, and a refined grid is therefore needed near $K=0$.
(5)~On the real axis Hermiticity preservation, Eq.~\eqref{eq:Jrule}, gives $\nB(0^-)=\nK(0^-)=W_0$. Twisting both indices moves $K$ by two steps and gives $\nC(0^-)=2W_0$, while a twist of the ket forward and the bra backward is a gauge transformation that leaves $K$ fixed and gives zero.

\emph{The case $l=2$.} The ring is still invariant under translation by one unit cell, so $K$ remains conserved, but each jump couples the same sublattice in neighboring cells, and the closed form~\eqref{eq:LK} does not apply as written. More importantly, $W_0$ counts the joint displacement of ket and bra, Eq.~\eqref{eq:nuW0}, and for $l=2$ it vanishes by Eqs.~\eqref{eq:Jrule} and \eqref{eq:Crule}. The bipolar behavior appears only at complex reference points and only when the two indices are twisted separately. For real $\Lambda$, $\det[\Lv_{\mathrm K}(\theta)-\Lambda]$ is real at every $\theta$, and the windings at complex references come in conjugate pairs of opposite sign (Sec.~\ref{supp:sym}).
\section{Symmetry and balance}
\label{supp:sym}

The symmetry statements of the main text rest on two ingredients. The first is the antiunitary map $\mathcal{A}$ of Eq.~\eqref{eq:sym}, assembled from the mirror $P$ ($c_i\to c_{N-1-i}$), the sublattice sign $\Gamma$ ($c_i\to(-1)^ic_i$), and complex conjugation in the site basis. The second is Hermiticity preservation, which every Lindbladian has. We first derive the balance relations that follow from a single jump, and then show how these maps act on the windings and on the steady state.

\subsection{Single-jump identities and balance relations}
\label{supp:balance}

Write the jump on the bond $(j,j+l)$ as $O=u^\dagger v$, with $u^\dagger=c_j^\dagger+z\,c_{j+l}^\dagger$, $v=c_j-z\,c_{j+l}$, and $\{u,u^\dagger\}=\{v,v^\dagger\}=2$. In the single-particle sector $O^\dagger O=2v^\dagger v$ and $OO^\dagger=2u^\dagger u$, so that
\begin{equation}
[O,O^\dagger]=2\big(u^\dagger u-v^\dagger v\big)=2(z+z^*)\big(c_j^\dagger c_{j+l}+c_{j+l}^\dagger c_j\big).
\end{equation}
Summed over bonds, $\Lv_\alpha(\id)=\gamma\sum_j[O_j,O_j^\dagger]$ is Eq.~\eqref{eq:unital}. It vanishes only for $u\propto v$, that is, for $z^*=-z$ or $\alpha=\pi/2+n\pi$, where the jump is Hermitian. Each part of Eq.~\eqref{eq:DT} is normal on its own, since $O^{D}$ is built from commuting number operators and $O^{T}O^{T\dagger}=O^{T\dagger}O^{T}=n_j+n_{j+l}$, and either part alone therefore gives a unital bath. For the full jump $[O,O^\dagger]=[O^{D},O^{T\dagger}]+[O^{T},O^{D\dagger}]$, and only the cross terms of Eq.~\eqref{eq:interference} remain.

\emph{Populations.} The bond adds $\gamma\langle O^\dagger n_jO-\tfrac12\{O^\dagger O,n_j\}\rangle$ to $\dot n_j$. With $O^\dagger n_jO=|\{c_j,u^\dagger\}|^2\,v^\dagger v=v^\dagger v$ and $\tfrac12\{O^\dagger O,n_j\}=\{v^\dagger v,n_j\}$,
\begin{equation}
\langle v^\dagger v\rangle=n_j+n_{j+l}-2\Real\big(z\langle c_j^\dagger c_{j+l}\rangle\big),\qquad
\langle\{v^\dagger v,n_j\}\rangle=2n_j-2\Real\big(z\langle c_j^\dagger c_{j+l}\rangle\big).
\end{equation}
The coherence drops out of the difference, which leaves $\gamma(n_{j+l}-n_j)$, Eq.~\eqref{eq:flatten}, and the same calculation applies to site $j+l$.

\emph{Current.} A bath bond $(a,b)$ carries the flow $\gamma(n_a-n_b)$ from $a$ to $b$, and the Hamiltonian changes the occupation of site $j$ at the rate $J_{j-1}-J_j$, with $J_j$ from Eq.~\eqref{eq:J}. Nothing crosses the cut to the left of site $0$, so adding the stationary site equations from the left sets the net flow across every cut to zero. The cut between $j$ and $j+1$ is crossed by one Hamiltonian bond together with one bath bond for $l=1$, or with the two bath bonds $(j-1,j+1)$ and $(j,j+2)$ for $l=2$, which gives the two lines of Eq.~\eqref{eq:balance}. On a ring the net flow is uniform instead. For $l=1$ the occupations are also uniform on each sublattice, the bath flows on the intracell and intercell bonds cancel, and the average over one cell is Eq.~\eqref{eq:Jring}.

\emph{Zero modes.} Because $[H,\Pi]=0$, only the jumps change $\langle\Pi\rangle$. In the single-particle sector $O_j^\dagger\Pi O_j=w_j\,v_j^\dagger v_j$, with $w_j=\langle\mathrm{vac}|u_j\Pi u_j^\dagger\vac=\sum_m|\{d_m,u_j^\dagger\}|^2$, and $\tfrac12\{O_j^\dagger O_j,\Pi\}=\{v_j^\dagger v_j,\Pi\}$. Together these give Eq.~\eqref{eq:zmbalance}.

\subsection{Windings under the mirror, Hermiticity, and \texorpdfstring{$\mathcal{A}$}{A}}
\label{supp:maps}

Three rules are used below. (R1)~A linear map $\mathcal{U}$ with $\mathcal{U}\Lv(\theta_{\mathrm K},\theta_{\mathrm B})\mathcal{U}^{-1}=\Lv(\theta'_{\mathrm K},\theta'_{\mathrm B})$ leaves the determinant unchanged, and with it the winding along corresponding loops. (R2)~An antilinear map conjugates the determinant, sends $\Lambda\to\Lambda^*$, and reverses the sense of rotation. (R3)~Running a loop backward reverses its winding.

\emph{Hermiticity.} Under $\mathcal{J}(\rho)=\rho^\dagger$ the operators acting on the ket and on the bra trade places; for the Hamiltonian part, $[-i(H_{\theta_{\mathrm K}}\rho^\dagger-\rho^\dagger H_{-\theta_{\mathrm B}})]^\dagger=-i(H_{-\theta_{\mathrm B}}\rho-\rho H_{\theta_{\mathrm K}})$, and the jump terms follow suit. Hence
\begin{equation}
\mathcal{J}\,\Lv(\theta_{\mathrm K},\theta_{\mathrm B})\,\mathcal{J}^{-1}=\Lv(-\theta_{\mathrm B},-\theta_{\mathrm K}).
\end{equation}
The ket loop becomes the bra loop run backward, and (R2) with (R3) yields Eq.~\eqref{eq:Jrule} for any number-conserving Lindbladian.

\emph{Mirror.} $P$ maps the bonds that cross the boundary onto themselves with their direction reversed, so $PH_\theta P^{-1}=H_{-\theta}$, while $PO_j(\alpha)P^{-1}=-z^2\,O_{N-1-j-l}(-\alpha)$ and the factor $-z^2$ cancels between $O$ and $O^\dagger$. The mirror is thus a linear map from $\Lv_\alpha(\theta_{\mathrm K},\theta_{\mathrm B})$ to $\Lv_{-\alpha}(-\theta_{\mathrm K},-\theta_{\mathrm B})$, so that every winding is odd in $\alpha$ and vanishes at $\alpha=0,\pi$ for both ranges, which is the triviality criterion of Ref.~\cite{Hamanaka2023}.

\emph{Mirror-chiral symmetry.} The sublattice sign by itself reverses $H$, keeps $O^{D}$, and multiplies $O^{T}$ by $(-1)^l$. Its antiunitary completion is $\mathcal{A}(\rho)=P\Gamma\rho^*\Gamma P$. Conjugation acts as $(\alpha,H,\theta)\to(-\alpha,-H,-\theta)$, because it flips the factor $i$ in $-i[H,\cdot]$; $\Gamma$ then restores $H$ and, for $l=1$ only, shifts $\alpha$ by $\pi$; finally $P$ reverses $\alpha$ and $\theta$ once more. The result is
\begin{equation}
\mathcal{A}\,\Lv_\alpha(\theta_{\mathrm K},\theta_{\mathrm B})\,\mathcal{A}^{-1}=\begin{cases}\Lv_\alpha(\theta_{\mathrm K},\theta_{\mathrm B}), & l=2,\\ \Lv_{\alpha+\pi}(\theta_{\mathrm K},\theta_{\mathrm B}), & l=1,\end{cases}
\label{eq:Asupp}
\end{equation}
which reduces to Eq.~\eqref{eq:sym} at zero twist.

For $l=2$, (R2) gives $\nu(\Lambda^*)=-\nu(\Lambda)$ for every winding, and $\det[\Lv(\theta)-\Lambda]$ is real for real $\Lambda$. The product $\mathcal{J}\mathcal{A}$, $\rho\mapsto P\Gamma\rho^{T}\Gamma P$, is linear and sends $(\theta_{\mathrm K},\theta_{\mathrm B})\to(-\theta_{\mathrm B},-\theta_{\mathrm K})$. The ket loop turns into the bra loop run backward and the density loop into itself run backward, so (R1) and (R3) give $\nK=-\nB$ and $\nC=0$, Eq.~\eqref{eq:Crule}. For $l=1$, $\mathcal{A}$ links $\alpha$ to $\alpha+\pi$ instead; then $\nu_{\alpha+\pi}(\Lambda^*)=-\nu_\alpha(\Lambda)$, and on the real axis every winding reverses under $\alpha\to\alpha+\pi$, in agreement with Eq.~\eqref{eq:W0sym}.

Equation~\eqref{eq:Crule} follows from $\mathcal{A}$ alone, and it is therefore exact for a uniform chain and in the presence of a staggered mass $\sum_i(-1)^in_i$, which the mirror maps onto its negative. Weak breaking of $\mathcal{A}$, by on-site disorder or by hoppings without mirror symmetry, leaves the integer windings practically unchanged, as expected for a symmetry-protected relation~\cite{Kawabata2019,Okuma2020}. The orientation of the bonds is also relevant. When every bond is ordered with its A site first, $\alpha$ is reversed on the intercell bonds and the $l=1$ pile-up is nearly removed (the steady-state center of mass at $\alpha=\pi/4$, $N=50$, falls from $+0.85$ to $+0.09$ on the scale $[-1,1]$).

\subsection{Steady state, Hermitian points, and zero modes}
\label{supp:ss}

For a unique steady state, the mirror and Eq.~\eqref{eq:sym} imply
\begin{equation}
\rss(-\alpha)=P\rss(\alpha)P,\qquad
\rss(\alpha+\pi)=\mathcal{A}\big[\rss(\alpha)\big]\ \ (l=1),\qquad
\rss(\alpha)=\mathcal{A}\big[\rss(\alpha)\big]\ \ (l=2).
\label{eq:rhoSym}
\end{equation}
Since $\mathcal{A}[\rho]_{ii}=\rho_{N-1-i,N-1-i}$, the occupations obey $\rho_{ii}(-\alpha)=\rho_{ii}(\alpha+\pi)=\rho_{N-1-i,N-1-i}(\alpha)$ for $l=1$, whereas for $l=2$ they are mirror symmetric at every $\alpha$ and admit no directional pile-up. For $H=0$, conjugation alone maps $\Lv_\alpha$ onto $\Lv_{-\alpha}$; combined with the mirror this gives $\rss^*=P\rss P$, a mirror-symmetric density at every $\alpha$. In the energy basis, the eigenmodes $d_E=\sum_i\phi_E(i)c_i$ of the real $H$ have real amplitudes, and for a nondegenerate spectrum $P\Gamma$ maps $d_E$ onto $\pm d_{-E}$. The population of $d_E$ in $\mathcal{A}[\rho]$ is therefore that of $d_{-E}$ in $\rho$, which together with Eq.~\eqref{eq:rhoSym} is Eq.~\eqref{eq:pE}.

\emph{Hermitian points.} At $\alpha_H=\pi/2+n\pi$ the dissipator is self-adjoint and $\Lv^\dagger=i[H,\cdot]+\mathcal{D}$. Conjugation takes $\Lv^\dagger_{\alpha_H}$ to $\Lv_{-\alpha_H}$, and the mirror takes this back to $\Lv_{\alpha_H}$, so that $\Lv^\dagger=\mathcal{S}^{-1}\Lv\,\mathcal{S}$ with the antiunitary $\mathcal{S}(X)=PX^*P$. This is Eq.~\eqref{eq:mirrorDB}. A right eigenmode $\rho_R$ with eigenvalue $\lambda$ thus has the left partner $\rho_L\propto P\rho_R^*P$, for both ranges. Detailed balance at infinite temperature, $\Lv^\dagger(X)=[\Lv(X^*)]^*$, would instead give $\rho_L\propto\rho_R^*$, with both at the same boundary~\cite{Haga2021}, whereas the mirror in $\mathcal{S}$ places them at opposite boundaries.

\emph{Zero modes.} $\Pi$ is invariant under $P$, $\Gamma$, and conjugation. The pair consists of $d_L$, on sublattice A at the left edge, and $d_R$, on sublattice B at the right edge, which $P\Gamma$ exchanges up to sign. With $p_{L,R}=\langle d_{L,R}^\dagger d_{L,R}\rangle$,
\begin{equation}
p_L(-\alpha)=p_R(\alpha),\qquad p_L(\alpha+\pi)=p_R(\alpha)\ \ (l=1),\qquad p_L(\alpha)=p_R(\alpha)\ \ (l=2),
\end{equation}
and $\PZM=2/N$ at $\alpha=\pi/2+n\pi$. The splitting within the pair is exponentially small in $N$, so the occupation of an individual member depends on the basis chosen inside the pair; we therefore use the pair weight $\PZM=\langle\Pi\rangle$ and its edge-resolved parts $p_{L,R}$.

For next-nearest-neighbor bonds the two ends are equivalent at every $\alpha$ and $N$. The pair is favored over equipartition for $\pi/2<\alpha<3\pi/2$, by up to an order of magnitude near $\alpha\approx0.9\pi$, yet its share of the particle shrinks as the chain grows. At $\alpha=\pi$ the jump on $(j,j+2)$ empties $\psi_j+\psi_{j+2}$ and so drains the sign-alternating zero mode at the size-independent rate $2\gamma(1-t_1/t_2)^2$, while bulk states with the same pattern become invisible to the bath at large $N$. No fixed share therefore survives, and the kink is absent. This favoring is unrelated to the bipolar skin effect of coherences, since the zero-mode weight is mirror symmetric about $\alpha=\pi$, where the bipolar imprint changes sign.

\section{Vacuum sector, corner rule, and skin modes}
\label{supp:vacuum}

\begin{figure*}[t]
\includegraphics[width=0.96\textwidth]{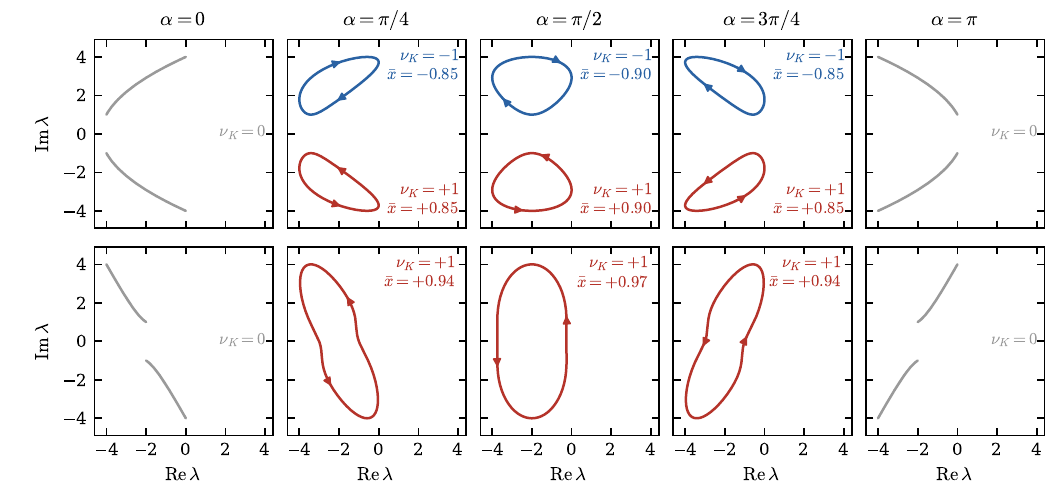}
\caption{Vacuum block of the full Liouvillian on a 24-site twisted ring, top row $l=2$, bottom row $l=1$. Red: counterclockwise loops ($\nK=+1$, right edge); blue: clockwise ($\nK=-1$, left edge); gray: arcs of zero area ($\nK=0$). Arrows give the sense of rotation as $\theta$ increases, and $\bar x\in[-1,1]$ is the mean position of the open-chain modes ($N=60$) belonging to each loop. $t_1=1.5$, $t_2=2.5$, $\gamma=1$.}
\label{figS:vac}
\end{figure*}

\begin{figure*}[t]
\includegraphics[width=0.96\textwidth]{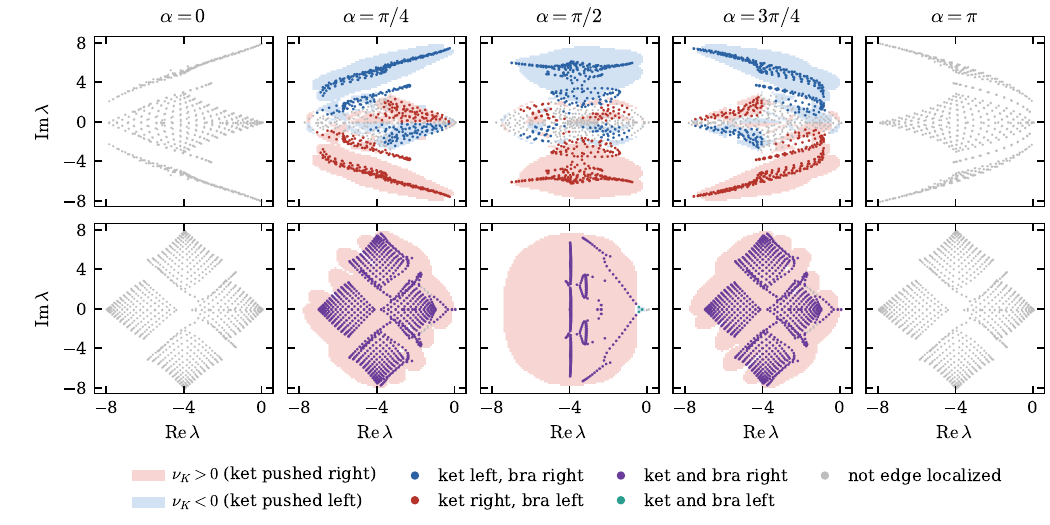}
\caption{Density block on the open chain ($N=30$), top row $l=2$, bottom row $l=1$. Points are the eigenvalues, colored by the corner in which the eigenmode localizes; shading is the sign of the ket winding $\nK(\Lambda)$ on a 12-site twisted ring. For $l=2$ every edge-localized mode sits in a crossed corner, in equal numbers; for $l=1$ every one sits in a same-edge corner. The few teal points near $\lambda=0$ at $l=1$, $\alpha=\pi/2$ lie in a small pocket with $\nK<0$ and sit at the left edge. $t_1=1.5$, $t_2=2.5$, $\gamma=1$.}
\label{figS:corner}
\end{figure*}

We explain the vacuum-sector loops of Figs.~\ref{fig:l1}(c), \ref{fig:l1}(d), \ref{fig:l2}(c), and \ref{fig:l2}(d), why they are weaker than Eq.~\eqref{eq:Crule}, how the shading and the corner rule were obtained, and why the edge-localized eigenmodes behave as skin modes. Figures~\ref{figS:vac} and \ref{figS:corner} extend the main-text panels to five phases, for both ranges.

\subsection{Blocks and vacuum loops}
\label{supp:loops}

On the space spanned by $\vac$ and $c_i^\dagger\vac$, group the rows and columns of $\rho$ by particle number. Because $H$ and every $O_j$ conserve particle number, each term of Eq.~\eqref{eq:lindblad} leaves the particle numbers of the row and of the column unchanged, so the blocks (ket number$\,|\,$bra number) $=(0|0)$, $(1|0)$, $(0|1)$, $(1|1)$ evolve independently.

\emph{The $(1|0)$ block.} Its elements are $\rho_{i,\mathrm{vac}}=\langle\mathrm{vac}|c_i\rho\vac$. Going through Eq.~\eqref{eq:lindblad}, we find that $H\rho$ acts on the ket and is kept, $\rho H$ vanishes because $H\vac=0$, $O_j\rho O_j^\dagger$ vanishes because $O_j^\dagger\vac=0$, and $O_j^\dagger O_j\rho$ is kept while $\rho O_j^\dagger O_j$ vanishes. What remains is
\begin{equation}
\frac{d}{dt}\rho_{\cdot,\mathrm{vac}}=G\,\rho_{\cdot,\mathrm{vac}},\qquad G=-iH-\frac{\gamma}{2}\sum_jO_j^\dagger O_j ,
\end{equation}
a matrix equation for $N$ numbers. The jump term therefore does not contribute in this block.

\emph{The $(1|1)$ block.} Writing $O_j=\sum_{ab}c_a^\dagger(o_j)_{ab}c_b$ and sorting terms by the index they act on,
\begin{equation}
\Lv_{11}=G\otimes\id+\id\otimes\bar G+\gamma\sum_jo_j\otimes\bar o_j .
\end{equation}
Without the last term every eigenvalue would be $\lambda_{\mathrm{row}}+\lambda_{\mathrm{column}}^*$, i.e., $N$ shifted copies of each vacuum loop, so the density-block spectra are dense. The jump term modifies them only through entries with $|i-j|\le l$.

\emph{Meaning.} The $(1|1)$ block describes the particle itself, and the $(1|0)$ block describes its amplitude. For qubits, $\rho_{i,\mathrm{vac}}=\langle\sigma_i^-\rangle$, while for fermions, where parity forbids it as a coherence of the state, the same block governs the propagator ``inject at $j$, find at $i$'' through the quantum regression theorem~\cite{Breuer2002}.

\emph{Vacuum loops.} Twisting the single index of the $(1|0)$ block on a 24-site ring moves its $N$ eigenvalues along one closed loop per band (Fig.~\ref{figS:vac}). For $l=2$ the two loops run in opposite senses for every $0<\alpha<\pi$. The loop above the real axis has $\nK=-1$ and its open-chain modes sit at the left edge, whereas the loop below has $\nK=+1$ and its modes sit at the right edge. The two loops are exact mirror images of each other, and $|\bar x|$ is largest at the Hermitian point. The mean position of the open-chain modes moves closer to the edges as $N$ grows. For $l=1$ there is one orientation, $\nK=+1$, with all modes at the right edge; the two parts of the loop merge into a single loop crossing the real axis for $\alpha$ between about $\pi/4$ and $3\pi/4$. Both ranges collapse to arcs at $\alpha=0,\pi$.

\emph{Why the vacuum loops are weaker than Eq.~\eqref{eq:Crule}.} A real bipartite $H$ with jumps that preserve chirality already gives $\bar G(\theta)=\Gamma G(\theta)^{T}\Gamma$. The spectrum of $G(\theta)$ then equals its complex conjugate, and $\nK(\Lambda^*)=-\nK(\Lambda)$, i.e., the two loops have opposite windings. This requires neither the mirror part of $\mathcal{A}$ nor the jump term. For hoppings without mirror symmetry the vacuum loops stay exactly paired, while the full-Liouvillian relation~\eqref{eq:Crule} no longer holds exactly. 

\subsection{Ket-winding shading and the transient coherence}
\label{supp:corner}

The shading in Figs.~\ref{fig:l1}(b), \ref{fig:l2}(b), and \ref{figS:corner} is $\sgn\nK(\Lambda)$ evaluated on a grid of reference points $\Lambda$ covering the density-block spectrum of a 12-site twisted ring. For an open-chain eigenmode $\rho_{ij}$ we define the ket and bra centers
\begin{equation}
\bar x_{\mathrm K}=\frac{\sum_{ij}x_i|\rho_{ij}|^2}{\sum_{ij}|\rho_{ij}|^2},\qquad
\bar x_{\mathrm B}=\frac{\sum_{ij}x_j|\rho_{ij}|^2}{\sum_{ij}|\rho_{ij}|^2},
\end{equation}
with $x_i=2i/(N-1)-1\in[-1,1]$, and call the mode edge localized if $|\bar x_{\mathrm K}|$ and $|\bar x_{\mathrm B}|$ both exceed $0.2$; its corner is $[\sgn\bar x_{\mathrm K},\sgn\bar x_{\mathrm B}]$. The corner rule predicts the ket side $\sgn\nK(\lambda)$ and, through Eqs.~\eqref{eq:Jrule} and \eqref{eq:Crule}, the bra side.

For $l=2$ only crossed corners occur, in equal numbers (Fig.~\ref{figS:corner}, top row), while for $l=1$ every localized mode sits in a same-edge corner (bottom row). The rule holds equally in the trivial phase, and its rare failures lie on the boundaries between regions of opposite $\nK$. For $l=1$ the sign of $\nK$ is uniform except at the Hermitian point, where it turns negative in a small pocket close to $\lambda=0$; the few modes inside it sit at the left edge, as the rule requires, while the slowest decaying mode stays extended.

To check that the localized modes are skin modes, we add a weak hop between sites $0$ and $N-1$. It shifts their eigenvalues by an amount that grows exponentially with $N$, and faster still for $l=1$, whereas at $\alpha=0$, where no winding is present, the shift decreases with $N$. The left eigenvector of each localized $l=2$ mode occupies the mirror corner of its right partner, as required for skin modes~\cite{Haga2021}, and the anticorrelation between ket and bra positions strengthens with size; for $l=1$ the two positions are instead positively correlated.

The bipolar structure is nevertheless transient. A bulk coherence, isolated by subtracting the full evolutions of $(c_{j_0}^\dagger\pm c_{j_0+1}^\dagger)\vac/\sqrt2$, spreads to both ends equally for $l=2$, to the right end only for $l=1$, and to neither at $\alpha=0$, and in every case it eventually decays. The $l=2$ steady state retains only a negligible end-to-end coherence, and what survives is a mirror-symmetric enhancement of short-range coherence at both edges, without a directional skin effect of the density.

\end{document}